\documentclass[11pt,oneside]{article}
\usepackage{a4wide}
\usepackage{mathrsfs}
\usepackage{latexsym,bm}
\usepackage{CJK}
\usepackage{graphicx}
\usepackage{indentfirst}
\usepackage{slashed}
\usepackage{amsmath}
\usepackage{amssymb}
\usepackage{color}
\usepackage{hyperref}
\usepackage{epsfig}
\hypersetup{CJKbookmarks=true}
\usepackage[titletoc]{appendix}
\usepackage{multirow}%
\usepackage{rotating}
\usepackage{epstopdf}
\usepackage{cite}
\usepackage{extarrows}
\usepackage[top=2.9cm,bottom=2.5cm,left=2.8cm,right=3cm]{geometry}
\usepackage{tabularx}

\newcommand{\be}{\begin{equation}} \newcommand{\ee}{\end{equation}}
\newcommand{\ba}{\left(\begin{array}{c}}
\newcommand{\ea}{\end{array}\right)}
\newcommand{\bea}{\begin{eqnarray}} \newcommand{\eea}{\end{eqnarray}}

\newcommand{\yao}{\color{blue}}

\newcommand{\al}{&\!\!\!\!}

\newcommand{\ds}{D_{s0}^*(2317)}

\newcommand{\bma}{\left(\begin{matrix}}
\newcommand{\ema}{\end{matrix}\right)}
\newcommand{\bqa}{\begin{eqnarray}}
\newcommand{\eqa}{\end{eqnarray}}
\newcommand{\bqaa}{\begin{eqnarray*}}
\newcommand{\eqaa}{\end{eqnarray*}}
\newcommand{\mL}{\mathcal{L}}
\newcommand{\cO}{\mathcal{O}}
\newcommand{\bra}{\langle}
\newcommand{\ket}{\rangle}
\newcommand{\nn}{\nonumber}
\newcommand{\mD}{\mathcal{D}}
\newcommand{\mP}{\mathcal{P}}
\newcommand{\mV}{\mathcal{V}}

\begin{document}
\thispagestyle{empty}
\title{ 
\Large \bf 
New insights into the $D^{*}_{s0}(2317)$ and other charm scalar mesons }
\author{\small Zhi-Hui Guo$^{a,b}$\thanks{zhguo@mail.hebtu.edu.cn}, Ulf-G. Mei{\ss}ner$^{c,d}$\thanks{meissner@hiskp.uni-bonn.de}, De-Liang Yao$^{d}$\thanks{d.yao@fz-juelich.de}\\[1mm]
{ \small\it ${}^a$  Department of Physics, Hebei Normal University,  Shijiazhuang 050024, }\\
{\small\it People's Republic of China }\\
{\small\it    ${}^b$ State Key Laboratory of Theoretical Physics, Institute of
Theoretical Physics, CAS, 
}\\
{\small\it Beijing 100190, People's Republic of China}
\\
{\small\it  $^c$Helmholtz-Institut f\"ur Strahlen- und Kernphysik and
Bethe Center for Theoretical Physics,}\\
{\small\it Universit\"at Bonn, D--53115
Bonn, Germany}\\
{\small\it  $^d$Institute for Advanced Simulation, Institut f{\"u}r
Kernphysik and J\"ulich Center for Hadron Physics,}\\ 
{\small\it Forschungszentrum 
J{\"u}lich, D-52425 J{\"u}lich, Germany}
}
\date{}

%
\maketitle
\begin{abstract}
Through the scattering of light-pseudoscalar mesons ($\pi,K,\eta,\eta'$) off charmed mesons ($D, D_s$), 
we study the $D^{*}_{s0}(2317)$ state and other relevant charm scalar mesons in a unitarized chiral 
effective field theory approach. 
We investigate the charm scalar meson poles with different strangeness~($S$) and isospin~($I$) 
quantum numbers as well as their corresponding residues, which provide the coupling strengths of the 
charm scalar mesons. Both the light-quark mass and $N_C$ dependences of the pole positions of 
the $D^{*}_{s0}(2317)$ and the poles with $(S,I)=(0,1/2)$ 
are analyzed in detail in this work. Interestingly we observe quite similar pion mass trajectories 
for the resonance pole at around 2.1~GeV with $(S,I)=(0,1/2)$ to those of the $f_0(500)$ given in 
the literature. 
When increasing the values of $N_C$ we find that a bound state and a virtual state in the 
$(S,I)=(1,0)$ channel asymmetrically approach the $D K$ threshold 
for $N_C<6$, and they meet at this threshold at $N_C=6$. When $N_C>6$, the bound and virtual 
states move into the complex plane on the second Riemann sheet 
and become a symmetric pair of resonance poles. For large enough values of $N_C$, neither 
the $\ds$ pole nor the poles with $(S,I)=(0,1/2)$ tend to fall down to the real axis, 
indicating that they do not behave like a standard quark-antiquark meson at large $N_C$. 
\end{abstract}
{\small PACS numbers: 12.39.Fe, 13.75.Lb, 14.40.Lb}

\newpage

\section{Introduction}

The discovery of the scalar charmed-strange meson $\ds$~\cite{babar03,cleo03,belle03} has triggered 
intensive studies in hadron physics. 
With its mass just about 40~MeV below the $DK$ threshold, it is one of the few precious 
examples of a loosely bound state observed in the meson sector. 
The scattering of the charmed and light pseudoscalar mesons provides a powerful tool to probe 
the inner structure of $\ds$. 
Recently much progress from  lattice simulations focusing on the $\ds$ state has been made 
investigating the previously mentioned scattering 
processes~\cite{Liu:2012zya,Mohler:2012na,Mohler:2013rwa,Lang:2014yfa}. The pion masses 
used in these lattice simulations are 
still large compared to the physical value, so the chiral extrapolations are necessary to obtain 
physical predictions from the lattice data. 
Chiral perturbation theory~($\chi$PT), as an effective field theory of low energy QCD, has 
been demonstrated to 
be a successful and reliable tool to perform the chiral extrapolations in many lattice 
simulations~\cite{Aoki:2013ldr}. 

{In traditional heavy meson $\chi$PT (HM$\chi$PT), proposed by Refs~\cite{Burdman:1992gh,Wise:1992hn,Yan:1992gz}, not only is the chiral symmetry encoded but also the heavy-quark symmetry is complemented so as to study heavy-light systems. Due to the incorporation of heavy-quark symmetry, the interactions are independent of heavy-quark spin and flavor, which imply that the scalar $D$ and vector $D^\ast$ should be treated as counterparts to form, for instance, a spin doublet. However, following 
Ref.~\cite{Guo:2009ct}, a covariant HM$\chi$PT, in combination only with the chiral symmetry, will be imposed to investigate the interactions between the charmed $D$ and pseudo Nambu-Goldstone bosons (pNGBs) in this work. In covariant $\chi$PT, the necessity of the inclusion of the $D^\ast$ relies on the size of its contributions. In our case,  only the $0^+$ charmed states involved in the $S$-wave interactions will be studied and the vector $D^\ast$ only contributes via $u$-channel exchange, which provides a negligible contribution to the interacting potential at tree level~\cite{Geng:2013vwa}. 
Therefore we consider it is a good approximation to exclude the vector $D^\ast$ mesons when focusing only on the $S$-wave charmed $D$ meson and pNGB scattering.} 
Within the framework of covariant heavy meson $\chi$PT by taking the $D_{(s)}$ and the octet of pNGBs as the dynamical fields, 
the leading order~(LO)~\cite{Lutz:2003ac,Hofmann:2003je,Guo:2006fu}, next-to-leading order 
(NLO)~\cite{Guo:2009ct,Cleven:2010aw,Wang:2012bu,Geng:2013vwa,Altenbuchinger:2013gaa} 
and next-to-next-to-leading order (NNLO)~\cite{Geng:2010vw,Yao:2015qia} calculations have been performed to 
investigate the properties of $\ds$. 
The pole trajectories with varying up/down- and strange-quark masses are useful 
quantities to further probe the inner structure of the hadrons, e.g. the $M_\pi$ 
and $M_K$ dependences of $\ds$ have been extensively 
studied in previous Refs~\cite{Cleven:2010aw,Wang:2012bu,Geng:2013vwa} and similar 
trajectories for the light-flavor resonances  
$f_0(500)$/$\sigma$ and $\rho(770)$ are predicted in Refs.~\cite{Hanhart:2008mx,Hanhart:2014ssa}. 
Another kind of pole 
trajectories with varying $N_C$, with $N_C$ the number of colors in QCD, has been  demonstrated 
to be useful to discriminate different 
explanations for light-flavor resonances~\cite{Sun:2005uk,Pelaez:2006nj,Dai:2011bs,Guo:2011pa,Guo:2012yt}, such as for $f_0(500)$, $\rho(770)$, etc. 
However, the study of the $N_C$ trajectories of the $\ds$ state is still 
lacking. One of the key novelties of the present work is to 
fill this gap. The $N_C$ trajectories for other pertinent charm scalar mesons resulting 
from the $D_{(s)}$ and pNGBs scattering  shall be investigated as well. 
We note that finite-volume corrections for this type of scattering processes have recently been
worked out, see Refs.~\cite{Torres:2014vna,Agadjanov:2014ana}.

We point out that in order to study the $N_C$ trajectories of the poles from the $D_{(s)}$ 
and pNGBs scattering, it is inappropriate 
to only consider the $SU(3)$ octet of pNGBs, i.e. $\pi, K, \eta_8$. The singlet $\eta_0$ becomes 
a relevant degree of freedom (d.o.f) when discussing large $N_C$. The reason is that the 
QCD $U(1)_A$ anomaly, which is responsible for the 
massive $\eta_0$ in the physical case at $N_C=3$, is $1/N_C$ suppressed in the large $N_C$ limit. 
As a result, the QCD spectrum is the 
pNGB nonet at low energy in the chiral and large $N_C$ limits~\cite{ua1nc0,ua1nc1,ua1nc2,Manohar:1998xv}, 
rather than the $SU(3)$ octet resulting in the chiral limit. 
Therefore it is necessary for us to generalize the previous discussions on the charmed mesons 
$D_{(s)}$ and the pNGB octet scattering, 
such as those in 
Refs.~\cite{Lutz:2003ac,Hofmann:2003je,Guo:2006fu,Guo:2009ct,Cleven:2010aw,Wang:2012bu,Geng:2013vwa,Yao:2015qia}, 
to the processes involving the singlet $\eta_0$. After the inclusion of the singlet $\eta_0$, 
not only the QCD spectrum is completed in the large $N_C$ limit, but also we can have a more 
realistic description for  the physical $\eta$ meson at $N_C=3$, since the physical $\eta$ and 
$\eta'$ mesons are mixtures of $\eta_8$ and $\eta_0$. While in the 
standard $SU(3)$ $\chi$PT, the $\eta_8$ is typically identified with the physical 
$\eta$ state~\cite{Gasser:1984gg} and $\eta-\eta'$ mixing effects appear in certain LECs, in particular in
$L_7$. Within the $\chi$PT framework with explicit  light 
pseudoscalar octet plus singlet $\pi, K, \eta, \eta'$ (recognized as large $N_C$ or $U(3)$ $\chi$PT~\cite{Kaiser:2000gs}) 
and the charmed mesons $D_{(s)}$, we give an updated discussion by taking into account the 
lattice simulations and  
pay special attention to the pole spectra in the scattering of light pseudoscalar and charmed mesons. 

The paper is organized as follows. We first calculate all of the amplitudes for 
the charmed meson ($D$, $D_{s}$) and light pseudoscalars ($\pi, K, \eta, \eta'$) scattering 
processes in  $U(3)$ $\chi$PT, 
and then we use a simple recipe to unitarize the perturbative scattering amplitudes. This will 
be the main focus of Sect.~\ref{sect.chptamp}. 
The fits to the lattice simulation data will be presented in Sect.~\ref{sect.fit}, where we 
determine all of the free parameters. 
The pole contents, the corresponding residues and the pole trajectories with varying 
light-quark masses and $N_C$ will be analyzed in detail in Sect.~\ref{sect.pole}. 
Finally, we give a summary and conclude in Sect.~\ref{sect.concl}.

\section{Chiral amplitudes and their unitarization}\label{sect.chptamp}

Due to inclusion of the singlet $\eta_0$, with its mass squared $M_0^2$ remaining massive 
even in the chiral limit and behaving  
as $1/N_C$ for $N_C \to \infty$, one needs to introduce the $1/N_C$ expansion together with 
conventional chiral expansion 
consisting of soft momentum squared $p^2$ and light-quark masses $m_q$, in order to strictly 
establish a consistent  power-counting~\cite{HerreraSiklody:1996pm,Kaiser:2000gs}. The 
simultaneous triple expansions on 
the soft momentum squared, light-quark masses and $1/N_C$, i.e. $\cO(\delta) \sim 
\cO(p^2) \sim \cO(m_q) \sim \cO(1/N_C)$, 
will change the hierarchy of some operators assigned by the conventional chiral power counting. 
Typical examples are 
$L_4 \bra u_\mu u^\mu \ket \bra \chi_+ \ket$ and $L_5 \bra u_\mu u^\mu \chi_+ \ket$ from the light 
flavor $SU(3)$ $\chi$PT. 
Although the operators accompanying $L_4$ and $L_5$ belong to the same order in the $SU(3)$ case, 
their orders are changed  in the triple-expansion scheme when the singlet $\eta_0$ is 
introduced~\cite{Kaiser:2000gs}, since they have different orders in $1/N_C$. 
Compared to the $SU(3)$ octet case, new operators in the $U(3)$ case with arbitrary 
number of the building block $X=\bra \ln \det U \ket \propto \eta_0$ term 
can appear~\cite{HerreraSiklody:1996pm}, 
with $U$ the matrix of the pNGBs. However, from  large $N_C$ argument, one more $X$ field 
introduces one more order of the $1/N_C$ suppression. Therefore, the new operators with extra 
terms of the $X$ field are more  suppressed by $1/N_C$. On the other hand, from the practical 
point of view, the operators with the $X$ term exclusively contribute to the processes 
involving $\eta$ and $\eta'$ mesons, which are only relevant to the coupled channels of the 
$D_{(s)}$ and pNGBs scattering.  While the present lattice simulations in the coupled channel 
case are still rare and bear large uncertainties, as shown in later discussions, 
it is still not possible to make sensible determinations of the LECs accompanied by the new 
operators with the $X$ term.  Moreover, since the singlet $\eta_0$ predominantly contributes to 
the physical $\eta'$ state, the contributions from the new operators 
with an $X$ term will mainly enter the channels involving the $\eta'$ meson, which have thresholds 
that are obviously higher than in the other channels. 
This further indicates the irrelevance of the operators with an $X$ term in the present work. 
So we shall stop including any operator beyond the NLO discussion in 
Refs.~\cite{Guo:2009ct,Cleven:2010aw,Wang:2012bu,Geng:2013vwa}, 
which also enables us to make clear comparisons with the previous results.

The LO Lagrangian that describes the interactions between the charmed-meson triplet 
$\mP=(D^0,D^+,D^+_{s})$ and pNGBs takes 
the form 
\bea\label{lolag}
\mathcal{L}^{(1)}_{\mP\phi}=\mathcal{D}_\mu \mP \mathcal{D}^\mu \mP^\dagger-\overline{M}_D^2 \mP \mP^\dagger\ ,
\eea
which also provides the kinetic terms for the charmed mesons. The quantity $\overline{M}_D$ stands 
for the mass of the charmed mesons in the chiral limit. 
The covariant derivative $\mD_\mu$ acting on the charmed mesons $\mP$ is given by 
\bea
\mathcal{D}_\mu \mP= \mP(\overset{\leftarrow}{\partial_\mu}+\Gamma_\mu^\dagger)\ ,
\qquad \mathcal{D}_\mu \mP^\dagger=(\partial_\mu+\Gamma_\mu) \mP^\dagger\ ,
\eea
where 
\begin{eqnarray}\label{defbb}
\Gamma_\mu  &=& \frac{1}{2}\bigg[ u^\dagger (\partial_\mu- i\,r_\mu) u
+ u (\partial_\mu- i\,\ell_\mu) u^\dagger \bigg]\,, \qquad 
u^2 = U = e^{i\frac{ \sqrt2\Phi}{ F}} \,, \nn \\
\Phi &=&   \bma
      \frac{1}{\sqrt{2}}\pi^0+\frac{1}{\sqrt{6}}\eta_8+\frac{1}{\sqrt{3}}\eta_0 & \pi^+ &K^+ \\
      \pi^- & -\frac{1}{\sqrt{2}}\pi^0+\frac{1}{\sqrt{6}}\eta_8+\frac{1}{\sqrt{3}}\eta_0 &K^0\\
      K^-& \overline{K}^0&\frac{-2}{\sqrt{6}}\eta_8+\frac{1}{\sqrt{3}}\eta_0\\
   \ema\ .
\end{eqnarray}
In the above equations, $r_\mu$ and $\ell_\mu$ denote the right- and left-hand  external sources. 
$F$ denotes the weak decay constant of the pNGBs in the chiral and 
large $N_C$ limits, with the normalization $F_\pi=92.2$~MeV. 
A clear difference in the present discussion, compared to the previous ones 
in Refs.~\cite{Lutz:2003ac,Hofmann:2003je,Guo:2006fu,Guo:2009ct,Cleven:2010aw,Wang:2012bu,Geng:2013vwa,Yao:2015qia}, 
is the inclusion of the singlet $\eta_0$ in pNGB matrix given in Eq.~\eqref{defbb}. 

The NLO Lagrangian that generalizes Eq.~\eqref{lolag} reads~\cite{Guo:2009ct}
\bea\label{nlolag}
\mathcal{L}^{(2)}_{\mP\phi} &=& \mP \left(-h_0\langle\chi_+\rangle-h_1{\chi}_+
+ h_2\langle u_\mu u^\mu\rangle-h_3u_\mu u^\mu\right) {\mP}^\dag \nonumber\\
&& + \mathcal{D}_\mu \mP\left({h_4}\langle u_\mu
u^\nu\rangle-{h_5}\{u^\mu,u^\nu\}\right)\mathcal{D}_\nu {\mP}^\dag\,,
\eea
with 
\begin{eqnarray}\label{defbb2}
\chi_+ \,=\, u^\dagger  \chi u^\dagger  +  u \chi^\dagger  u \,, \qquad
u_\mu \, =\,
i \{ u^\dagger (\partial_\mu - i r_\mu) u\, -\, u(\partial_\mu - u\ell_\mu) u^\dagger\} \,, 
\end{eqnarray}
where $\chi=2 B (s + i p)$  includes the scalar ($s$) and pseudoscalar ($p$) external sources. 
The quantity $B$ is related 
to the light-quark condensate via $\bra 0| \bar{q}^i q^j|0\ket = -F^2 B\delta^{ij}$ at leading order.  
The light-quark masses in $\chi$PT are introduced by taking $(s + i p)$ =
diag($\hat{m},\hat{m},m_s$), 
with $\hat{m}$ the average up and down quark mass and $m_s$ the strange quark mass. Isospin 
violating effects will not  be considered throughout this work. Though in the strict triple 
expansion of $U(3)$ $\chi$PT the LECs in Eq.~\eqref{nlolag} 
belong to different orders, we simply quote them as the NLO ones as in 
Refs.~\cite{Guo:2009ct,Yao:2015qia}, since no other 
new operators with additional $X$ field will be considered and in this way it enables us to make 
clear comparisons with the results in the literature.

The relevant chiral Lagrangian with only the pNGB d.o.f is given 
by~\cite{ua1nc0,ua1nc1,ua1nc2,Kaiser:2000gs,HerreraSiklody:1996pm} 
\begin{eqnarray}\label{lolagphi}
\mL_\chi= \frac{ F^2}{4}\bra u_\mu u^\mu \ket+
\frac{F^2}{4}\bra \chi_+ \ket
+ \frac{F^2}{12}M_0^2 X^2 \,,
\end{eqnarray}
with $X=\ln(\det U)$. At leading order, the physical $\eta$ and $\eta^\prime$ can be obtained 
through the diagonalization of 
the $\eta_0$ and $\eta_8$ states in Eq.~\eqref{lolagphi}, and they are related by 
\bea
\ba \eta \\ \eta^\prime \\ \ea=
\bma
          c_\theta& -s_\theta \\
         s_\theta & c_\theta \\
\ema
\ba\eta_8 \\ \eta_0\\ \ea\ ,
\eea
with $s_\theta=\sin\theta$, $c_\theta=\cos\theta$ and $\theta$ the mixing angle. 
The masses of the $\eta$ and the $\eta'$ and their mixing angle $\theta$ can be expressed 
in terms of the parameters in Eq.~\eqref{lolagphi} 
\begin{eqnarray}
M_{\eta}^2 &=& \frac{M_0^2}{2} + M_K^2
- \frac{\sqrt{M_0^4 - \frac{4 M_0^2 \Delta^2}{3}+ 4 \Delta^4 }}{2} \,, \label{defmetab2}  \\
M_{\eta'}^2 &=& \frac{M_0^2}{2} + M_K^2
+ \frac{\sqrt{M_0^4 - \frac{4 M_0^2 \Delta^2}{3}+ 4 \Delta^4 }}{2} \,, \label{defmetaPb2}  \\
\sin{\theta} &=& -\left( \sqrt{1 +
\frac{ \big(3M_0^2 - 2\Delta^2 +\sqrt{9M_0^4-12 M_0^2 \Delta^2 +36 \Delta^4 } \big)^2}{32 \Delta^4} } ~\right )^{-1}\,,
\label{deftheta0}
\end{eqnarray}
with $\Delta^2 = M_K^2 - M_\pi^2$. We mention that it is enough for us to consider the LO chiral 
Lagrangian for the pNGBs
in Eq.~\eqref{lolagphi}. The LO square masses of the pion and the kaon are linearly 
dependent on the quark masses via 
\begin{equation}\label{mpimq}
 M_\pi^2 = 2 B \hat{m}\,, \qquad   M_K^2 = B (\hat{m} + m_s )\,.
\end{equation}

With the previous setup, it is straightforward to calculate the charmed mesons ($D$ and $D_s$) 
and pNGBs ($\pi, K, \eta$ and $\eta'$) 
scattering amplitudes, which can be categorized into seven different stes of quantum numbers
 characterized by strangeness ($S$) and isospin ($I$). 
The general expressions for the scattering amplitudes $D_1(p_1) + \phi_1(p_2) \to D_2 (p_3) + \phi_2(p_4)$ 
with definite strangeness and isospin can be written as 
\begin{equation}\label{eq:v}
V^{(S,I)}_{D_1\phi_1\to D_2\phi_2}(s,t,u) = \frac1{F^2} \bigg[\frac{C_{\rm LO}}{4}(s-u) - 4 C_0 h_0 +
2 C_1 h_1 - 2C_{24} H_{24}(s,t,u) + 2C_{35} H_{35}(s,t,u) 
\bigg]\,,  
\end{equation}
where $s,t,u$ are the standard Mandelstam variables, and $H_{24}(s,t,u)$ and $H_{35}(s,t,u)$ are  
\begin{eqnarray}
H_{24}(s,t,u) &=& 2 h_2 p_2\cdot p_4 + h_4 (p_1\cdot p_2 p_3\cdot p_4 +
p_1\cdot p_4 p_2\cdot p_3)\,, \\
H_{35}(s,t,u) &=& h_3 p_2\cdot p_4 + h_5
(p_1\cdot p_2 p_3\cdot p_4 + p_1\cdot p_4 p_2\cdot p_3)\,.
\end{eqnarray}
{The coefficients $C_i$ can be found in Table~\ref{tab:ci} and some explicit expressions of $C_i$ are relegated to the Appendix.}
 
Notice that we have additional channels for the $(S,I)=(1,0)$ and $(0,1/2)$ cases compared 
to the previous $SU(3)$ calculations~\cite{Guo:2009ct,Liu:2012zya,Geng:2013vwa,Wang:2012bu}  
and the scattering amplitudes involving the $\eta$ meson 
are also different from those references.  Nevertheless,  if we set the mixing angle $\theta$ 
to zero and use the Gell-Mann-Okubo mass relation 
$M_\eta^2= (4M_K^2-M_\pi^2)/3$, we explicitly verify that our results in Table~\ref{tab:ci} 
confirm the expressions in Refs.~\cite{Guo:2009ct,Liu:2012zya}. 
{ While comparing with the formulas in Ref.~\cite{Wang:2012bu}, our results do not agree 
with the item at the intersection of the column labeled 
by $C_1$ and the row labeled by $D_s\eta \to D_s\eta$. We also find that there are 
global sign differences between ours and 
those in Ref.~\cite{Geng:2013vwa} for the $D \pi \to D \eta$ and 
$D \pi \to D_s \bar{K}$ amplitudes in the case with $(S,I)=(0,1/2)$.

\begin{table}[htbp]
\centering
\begin{tabular}{|l c | c c c c c | }
\hline\hline
 $(S,I)$ & Channels & $C_{\rm LO}$ & $C_0$ & $C_1$ & $C_{24}$ & $C_{35}$ 
 \\
\hline
$(-1,0)$      & $D\bar{K}\to D\bar K$   & $-1$  & $M_K^2$ & $M_K^2$  & 1 & $-1$ 
\\
$(-1,1)$      & $D\bar{K}\to D\bar K$   & 1     & $M_K^2$ & $-M_K^2$ & 1 & 1 
\\
$(2,\frac12)$ & $D_sK\to D_sK$          & 1     & $M_K^2$ & $-M_K^2$ & 1 & 1 
\\
$(0,\frac32)$ & $D\pi\to D\pi$          & 1 & $M_\pi^2$ & $-M_\pi^2$ & 1 & 1  
\\
\hline
$(1,1)$       & $D_s\pi\to D_s\pi$      & 0   & $M_\pi^2$ & 0        & 1 & 0  
\\
              & $D K\to D K$            & 0     & $M_K^2$ & 0        & 1 & 0 
\\
              & $D K\to D_s\pi$   & 1 & 0 & $-(M_K^2+M_\pi^2)/2$ & 0    & 1
\\
\hline
$(1,0)$       & $D K\to D K$            & $-2$ & $M_K^2$ & $-2M_K^2$ & 1 & 2
\\
             & $D K\to D_s\eta$ & $-\sqrt{3} c_\theta$ & 0 &
        $ C_1^{1,0 \,\, D K\to D_s\eta} $ & 0 
        & $C_{35}^{1,0 \,\, D K\to D_s\eta}  $  
\\
              & $D_s\eta\to D_s\eta$ & 0 & $ C_0^{1,0 \,\, D_s\eta\to D_s\eta}$ 
              &  $C_1^{1,0 \,\, D_s\eta\to D_s\eta}$
             & 1 & $C_{35}^{1,0 \,\, D_s\eta\to D_s\eta}$ 
\\
             & $D K\to D_s\eta'$ & $-\sqrt{3} s_\theta$ & 0 &
        $ C_1^{1,0 \,\, D K\eta\to D_s\eta'} $ & 0 
        &  $ C_{35}^{1,0 \,\, D K\eta\to D_s\eta'}$ 
\\
              & $D_s\eta\to D_s\eta'$ & 0 & $ C_0^{1,0 \,\, D_s\eta\to D_s\eta'}$ 
              &  $C_1^{1,0 \,\, D_s\eta\to D_s\eta'}$
             & 0 & $C_{35}^{1,0 \,\, D_s\eta\to D_s\eta'}$
\\
              & $D_s\eta'\to D_s\eta'$ & 0 & $ C_0^{1,0 \,\, D_s\eta'\to D_s\eta'}$ 
              &  $C_1^{1,0 \,\, D_s\eta'\to D_s\eta'}$
             & 1 & $C_{35}^{1,0 \,\, D_s\eta'\to D_s\eta'}$ 
             
\\ 
\hline 
$(0,\frac12)$ & $D\pi\to D\pi$       & $-2$    & $M_\pi^2$ & $-M_\pi^2$ & 1 & 1
\\
             & $D\eta\to D\eta$     & 0 & $C_{0}^{0,\frac12\,\,D\eta\to D\eta}$ & $C_{1}^{0,\frac12\,\,D\eta\to D\eta}$& 1
             & $C_{35}^{0,\frac12\,\,D\eta\to D\eta}$  
\\
             & $D_s\bar K\to D_s\bar K$& $-1$& $M_K^2$& $-M_K^2$& 1 & 1 
\\ 
             & $D\eta\to D\pi$     & 0 & 0 & $M_\pi^2(\sqrt2 s_\theta- c_\theta)$ & 0 & $c_\theta-\sqrt2s_\theta$ 
\\
             & $D_s\bar K\to D\pi$ & $-\frac{\sqrt{6}}{2}$ & 0 &
             $-{\sqrt{6}}(M_K^2+M_\pi^2)/4$ & 0 & $\frac{\sqrt{6}}{2}$  
\\
             & $D_s\bar K\to D\eta$& $-\frac{\sqrt{6}}{2} c_\theta$ & 0 &
             $ C_{1}^{0,\frac12\,\,D_s K\to D\eta}$ & 0 & $ C_{35}^{0,\frac12\,\,D_s K\to D\eta}$ 
\\ 
             & $D\eta'\to D\pi$     & 0 & 0 & $-M_\pi^2(\sqrt2 c_\theta + s_\theta)$ & 0 & $s_\theta+\sqrt2 c_\theta$ 
\\
             & $D\eta\to D\eta'$     & 0 & $C_{0}^{0,\frac12\,\,D\eta\to D\eta'}$ & $C_{1}^{0,\frac12\,\,D\eta\to D\eta'}$& 0
             & $C_{35}^{0,\frac12\,\,D\eta\to D\eta'}$  
\\
             & $D_s\bar K\to D\eta'$     & $-\frac{\sqrt6}{2} s_\theta$ & 0 & $C_{1}^{0,\frac12\,\,D_s\bar K\to D\eta'}$& 0
             & $C_{35}^{0,\frac12\,\,D_s\bar K\to D\eta'}$ 
\\
             & $D\eta'\to D\eta'$     & 0 & $C_{0}^{0,\frac12\,\,D\eta'\to D\eta'}$ & $C_{1}^{0,\frac12\,\,D\eta'\to D\eta'}$& 1
             & $C_{35}^{0,\frac12\,\,D\eta'\to D\eta'}$  
\\
\hline\hline
\end{tabular}
\caption{\label{tab:ci}The coefficients in the scattering amplitudes 
$V^{(S,I)}_{D_1\phi_1\to D_2\phi_2}(s,t,u)$ in Eq.~\eqref{eq:v}. The
channels are labeled by strangeness ($S$) and isospin ($I$). For the coefficients not 
shown explicitly in this table, we give their 
expressions in the Appendix.} 
\end{table}

To continue the discussion, we perform the partial wave projections of the full amplitudes 
in Eq.~\eqref{eq:v} 
and the projection formula with definite angular momentum $J$ is given by
\bea\label{pwv}
\mathcal{V}_{J,\,D_1\phi_1\to D_2\phi_2}^{(S,I)}(s)
= \frac{1}{2}\int_{-1}^{+1}{\rm 
d}\cos\varphi\,P_J(\cos\varphi)\, V^{(S,I)}_{D_1\phi_1\to
D_2\phi_2}(s,t(s,\cos\varphi))\,,
\label{eq:pwp}
\eea
where $\varphi$ stands for the scattering angle between the incoming and outgoing particles 
in the center-of-mass frame,  
and the Mandelstam variable $t$ is given by 
\bea
t(s,\cos\varphi)\al=\al M_{D_1}^2+M_{D_2}^2-
\frac{1}{2s}\left(s+M_{D_1}^2-M_{\phi_1}^2\right)
\left(s+M_{D_2}^2-M_{\phi_2}^2\right) \nonumber\\
\al\al
-\frac{\cos\varphi}{2s}\sqrt{\lambda(s,M_{D_1}^2,M_{\phi_1}^2)
\lambda(s,M_{D_2}^2,M_{\phi_2}^2)}\ .
\label{eq:t}
\eea
with $\lambda(a,b,c)=a^2+b^2+c^2-2ab-2bc-2ac$  the K\"all\'{e}n function. 
We will be only interested in the $S$-wave projection in this work, i.e. the $J=0$ case. 
Therefore, the subscript $J$ labeling different partial wave amplitudes 
$\mathcal{V}_{J,\,D_1\phi_1\to D_2\phi_2}^{(S,I)}(s)$ will be dropped in later discussions for simplicity.

The perturbative scattering amplitudes at any finite order alone can not generate resonances 
or bound states.  Unitarity must be taken into account to study the resonances or bound states 
in the two-body scattering processes. 
We use the unitarization approach that has been extensively employed to discuss the $\ds$ 
previously in Refs.~\cite{Guo:2006fu,Guo:2009ct,Cleven:2010aw,Wang:2012bu,Geng:2013vwa}. This 
unitarization method was also 
used to study many other important phenomenons in hadron physics, such as the 
light-flavor meson resonances~\cite{Oller:1998zr,Guo:2011pa,Guo:2012yt}, 
the $\Lambda(1405)$~\cite{Oller:2000fj}, etc. 
The unitarized two-body scattering amplitude in this formalism is given by~\cite{Oller:2000fj} 
\begin{eqnarray} \label{defut}
 T(s) = \big[ 1 - \mV(s)\cdot g(s) \big]^{-1}\cdot \mV(s)\,,
\end{eqnarray}
where $\mV(s)$ stands for the partial wave amplitude in Eq.~\eqref{pwv} and for simplicity 
the super--  and subscripts have been dropped. The function $g(s)$ collects the unitarity cuts generated
 by the intermediate two-particle states in question. 
The loop function $g(s)$ is 
\bea
g(s)=i\int\frac{{\rm d}^4q}{(2\pi)^4}
\frac{1}{(q^2-M_{D}^2+i\epsilon)[(P-q)^2-M_{\phi}^2+i\epsilon ]}\ ,\qquad
s\equiv P^2\ \,,
\eea
which can be calculated in a once-subtracted dispersion relation or in dimensional regularization 
by replacing the divergence by a constant. The explicit form of $g(s)$ 
reads~\cite{Oller:1998zr}
\bea\label{gfunc}
g(s) \al=\al\frac{1}{16\pi^2}\bigg\{{a}(\mu)+\ln\frac{M_{D}^2}{\mu^2}
+\frac{s-M_{D}^2+M_{\phi}^2}{2s}\ln\frac{M_{\phi}^2}{M_{D}^2}\nonumber\\
\al\al+\frac{\sigma}{2s}\big[\ln(s-M_{\phi}^2+M_{D}^2+\sigma)-\ln(-s+M_{\phi}^2-M_{D}^2+\sigma)\nonumber\\
\al\al+\ln(s+M_{\phi}^2-M_{D}^2+\sigma)-\ln(-s-M_{\phi}^2+M_{D}^2+\sigma)\big]\bigg\}\ ,
\label{eq:g}
\eea
with $\sigma=\sqrt{\lambda(s,M_D^2,M_\phi^2)}$ and $\mu$ the regularization scale. 
The function $g(s)$ is independent of the scale $\mu$ and the explicit $\mu$-dependence in 
Eq.~\eqref{gfunc} is compensated by the  $\mu$-dependent subtraction constant $a(\mu)$. To be 
specific, we set $\mu=1$~GeV in the numerical discussions, in order to make 
a clear comparison with the previous works~\cite{Guo:2009ct,Yao:2015qia}. In this way, the value 
of the subtraction constant $a(\mu)$ determined in our fits correspond to its value at the scale $1$~GeV. 

The formalism for the unitarized partial-wave amplitude in Eq.~\eqref{defut} can be easily 
generalized to the coupled-channel case,  where the quantities $\mV(s)$ and $g(s)$ should be 
understood as matrices. The matrix elements for $\mV(s)$ can be calculated using 
Eq.~\eqref{pwv}. Further, $g(s)$ in the coupled-channel case becomes a diagonal matrix, with 
its matrix elements given by Eq.~\eqref{gfunc} 
with the pertinent values for $M_D$ and $M_\phi$.

\section{Fits to lattice simulation data}\label{sect.fit}

Up to now, there are  no experimental measurements on the charmed meson and pNGB scattering. 
On the other hand, lattice QCD  simulations are much advanced in this research 
field~\cite{Liu:2012zya,Mohler:2012na,Mohler:2013rwa,Lang:2014yfa} and can provide us with 
valuable information to constrain the interactions between the charmed mesons and the pNGBs. 
The useful information that lattice simulations provides are the $D \phi$ scattering lengths. 
In the channel with definite strangeness and isospin, the $S$-wave scattering lengths are 
related to the unitarized 
scattering amplitudes $T(s)$ in Eq.\eqref{defut} via 
 \bea
 a^{(S,I)}_{D\phi\to D\phi}=-\frac{1}{8\pi(M_D+M_{\phi})}T^{(S,I)}_{J=0}(s_{\rm thr})_{D\phi\to D\phi},
\qquad s_{\rm thr}=(M_D+M_\phi)^2\ .
 \eea
In order to make a comparison with the exsiting lattice-QCD data, the scattering lengths 
should be extrapolated to the unphysical quark masses via 
{
\bea\label{mdmpi}
 M_K &=& \sqrt{\overset{\circ\,\,}{M_K^2}+\frac{1}{2}M_\pi^2 }\ , \nn\\
 M_D &=& \sqrt{{\overset{\circ\,\,}{M_D^2}+2( 2h_0+h_1)M_\pi^2} }\ , \nn\\ 
 M_{D_s} &=& \sqrt{{\overset{\circ\,\,}{M_{D_s}^2}+4h_0M_\pi^2}} \ ,
\label{eq:chiralextra}
\eea 
}
which are obtained from Eqs.~\eqref{lolag}, \eqref{nlolag} and \eqref{lolagphi}. 
Here, $\overset{\circ\,\,}{M_K}$,  $\overset{\circ\,\,}{M_D}$ and  $\overset{\circ\,\,}{M_{D_s}}$
denote   the masses in the limit of $M_\pi^2~( \propto\hat{m} ) \to 0$, but with the strange 
quark mass $m_s \neq 0$ (the so-called two-flavor chiral limit). 
We mention that the quantities $\overset{\circ\,\,}{M_K}$,  $\overset{\circ\,\,}{M_D}$ and  
$\overset{\circ\,\,}{M_{D_s}}$  do not represent their values in the (three-flavor) chiral limit, 
since they are still dependent on $m_s$. Their forms read 
 \bea\label{mdms}
 {\overset{\circ\,\,}{M_K^2}} &=&  B m_s \, , \nn\\ 
  {\overset{\circ\,\,}{M_D^2}} &=& \overline{M}_D^2 +  4 h_0  {\overset{\circ\,\,}{M_K^2}}  \, ,  \nn\\ 
 {\overset{\circ\,\,}{M_{D_s}^2}} &=& \overline{M}_D^2 +  4 (h_0+h_1)  {\overset{\circ\,\,}{M_K^2}}   \, , 
\label{eq:chiralextra1}
 \eea 
where $\overline{M}_D$ is the chiral limit mass of the charmed mesons, see Eq.~\eqref{lolag}. 
Combining Eqs.~\eqref{mpimq}, \eqref{mdmpi} and \eqref{mdms}, one can easily perform the 
chiral extrapolations by varying the light quark mass $\hat{m}$ and the strange quark mass $m_s$.  
The corresponding light-quark mass dependences for $M_\eta$, $M_{\eta^\prime}$ and the mixing 
angle $\theta$ are obtained by combinig Eqs.~\eqref{mpimq}, \eqref{defmetab2}, \eqref{defmetaPb2} 
and \eqref{deftheta0}.  Compared to the previous works in 
Refs.~\cite{Guo:2009ct,Cleven:2010aw,Wang:2012bu}, we do not further make the expansion of $M_\pi$ 
inside the square roots in the right-hand-sides of Eq.~(\ref{eq:chiralextra}) in order to 
obtain $M_K, M_D$ and $M_{D_s}$. It is worthy noting that the present lattice simulations are performed with fixed strange- and 
charm-quark masses,  while only the up/down-quark masses are varied. In addition, the fixed 
strange and charm masses are usually set at (slightly) unphysical values in lattice QCD 
simulations and, furthermore, different configurations may adopt different values, which 
finally lead to different values for $\overset{\circ\,\,}{M_K}$,  $\overset{\circ\,\,}{M_D}$ 
and  $\overset{\circ\,\,}{M_{D_s}}$. In our case, the lattice QCD data are taken 
from two collaborations, e.g. Ref.~\cite{Liu:2012zya} and Refs.~\cite{Mohler:2012na,Mohler:2013rwa,Lang:2014yfa},  and the values of $\overset{\circ\,\,}{M_K}$,  $\overset{\circ\,\,}{M_D}$ 
and  $\overset{\circ\,\,}{M_{D_s}}$ corresponding to different lattice configurations, 
as well as their values extracted from the physical case, are listed in Table~\ref{tab:chirallimit} 
for easy comparison.

\begin{table}[htbp]
 \centering
\begin{tabular}{ c | c c c c}
\hline\hline
& Ref.~\cite{Liu:2012zya}& Ensemble (1)~\cite{Mohler:2012na,Mohler:2013rwa,Lang:2014yfa}& Ensemble (2)~\cite{Mohler:2012na,Mohler:2013rwa,Lang:2014yfa}& Physical
\\\hline
$\overset{\circ\,\,}{M_K}$~[MeV]&560.8&519.0&482.8&486.3
\\
$\overset{\circ\,\,}{M_D}$~[MeV]&1940&1538&1631&1862
\\
$\overset{\circ\,\,}{M_{D_s}}$~[MeV]&2058&1655&1731&1967
\\
\hline\hline
\end{tabular}
\caption{\label{tab:chirallimit} The masses defined in the limit $M_\pi \to 0$ in 
Eq.~\eqref{mdms}  for the chiral extrapolations.  }
\end{table}

 Next we  present our fit procedure. The unknown parameters are $F$, $M_0$, $h_{i=0,1,2,3,4,5}$ and 
the subtraction constants $a$.  Since we only work at leading order in the pNGB sector~\eqref{lolagphi}, 
it is justified to approximate the value of $F$ by the physical pion decay constant $F_\pi=92.2$~MeV. 
For $M_0$, we adopt the LO value $M_0=835.7$~MeV determined in Ref.~\cite{Guo:2015xva}, which 
 has taken into account the recent lattice simulation data on the $\eta$ and $\eta'$ 
masses~\cite{Michael:2013gka}. The LECs $h_0$ and $h_1$ are determined by the masses of the 
$D$ and $D_s$ mesons via Eq~\eqref{mdmpi} and $h_0$ by the slopes of lattice QCD data for the masses 
of the  $D$ and the $D_s$ while $h_1$ is fixed from the physical mass splitting between $D$ and $D_s$.  
The resulting numerical values can be found in the first two rows in Table~\ref{tab:LECs}.  

Further, redefinitions of the remaining LECs, i.e. $h_{i=2,3,4,5}$, are introduced in order to 
reduce the correlations during the fitting process, as done in Refs.~\cite{Liu:2012zya,Yao:2015qia}. 
These take the form 
 \bea\label{redfhi}
 h_{24}\equiv h_2+h_4^\prime\ , \quad h_{35}\equiv h_3+2\,h_5^\prime\ ,\quad 
h_4^\prime\equiv h_4 \hat{M}_D^2\ , \quad h_5^\prime\equiv h_5 \hat{M}_D^2\ ,
 \eea
with $\hat{M}_D\equiv(M_D^{\rm phys}+M_{D_s}^{\rm phys})/2$ the averaged physical masses of $D$ and 
$D_s$. The redefined LECs are dimensionless and used as fitting variables. 
A common subtraction constant $a$, occurring in the loop function in Eq.~\eqref{gfunc}, is 
adopted for all the possible channels~\footnote{In general, the subtraction constant $a(\mu)$ should be channel dependent. However, if distinct subtraction constants are adopted in different channels, our fits tend to be unstable. This indicates that the present lattice simulation data are still not precise enough to discriminate the individual subtraction constants in different channels. }. Then the redefined LECs 
$h_i$ in Eq.~\eqref{redfhi} and the subtraction constant $a$ are fitted to the lattice 
scattering lengths provided  in 
Refs.~\cite{Liu:2012zya,Mohler:2012na,Mohler:2013rwa,Lang:2014yfa}. 
We stress that our fits are done using directly the lattice QCD data for the various masses, 
rather than those derived from Eq.~\eqref{eq:chiralextra} together with the masses shown in 
Table~\ref{tab:chirallimit}. 
We mention that our treatment of the masses in the fits is the same as that in 
Ref.~\cite{Geng:2013vwa}, but different from the one in Ref.~\cite{Guo:2009ct}, which 
used the masses resulting from the chiral extrapolations in  Eq.~\eqref{eq:chiralextra}. 
For reference, the correlation coefficients between the fit parameters for Fit-6C from the CERN MINUIT minimization package are 
shown in Table~\ref{corcoe}. It is worthy noting that the correlations of the different lattice simulation data are not considered 
in our fits and hence the correlations of the fit parameters might be underestimated.

 \begin{table}[htbp]
 \centering
\begin{tabular}{ c r r r }
\hline\hline
LEC&Fit-6C& Fit-5C& Table V~\cite{Liu:2012zya}
\\ \hline
$h_{0}$&$0.033^\ast$&$0.033^\ast$&$0.014^\ast$
\\
$h_{1}$&$0.43^\ast$&$0.43^\ast$&$0.42^\ast$
\\
$h_{24}$&$-0.13_{-0.06}^{+0.05}$&$-0.13_{-0.06}^{+0.05}$&$-0.10_{-0.06}^{+0.05}$
\\
$h_{35}$&$0.23_{-0.06}^{+0.06}$&$0.24_{-0.12}^{+0.12}$&$0.25_{-0.13}^{+0.13}$
\\
$h_{4}^\prime$&$-0.21_{-0.27}^{+0.29}$&$-0.19_{-0.31}^{+0.32}$&$-0.32_{-0.34}^{+0.35}$
\\
$h_{5}^\prime$&$-1.78_{-0.19}^{+0.19}$&$-1.83_{-0.56}^{+0.57}$&$-1.88_{-0.61}^{+0.63}$
\\
${a}$($\lambda=1$~GeV)&$-1.88_{-0.06}^{+0.06}$&$-1.88_{-0.09}^{+0.07}$&$-1.88_{-0.09}^{+0.07}$
\\ \hline
$\chi^2/{\rm d.o.f}$&$\frac{12.27}{16-5}\cong1.12$&$\frac{12.26}{15-5}\cong1.23$&1.06
\\
\hline\hline
\end{tabular}
\caption{\label{tab:LECs} Fitting results of the parameters.  The fitting results labeled 
by Table V are taken from Ref.~\cite{Liu:2012zya} for comparison. 
Fit-6C and Fit-5C denote six- and five-channel fits, respectively. 
$h_0$ and $h_1$ are determined by the masses of the $D$ and the $D_s$, not by the 
scattering lengths.  See the text for details.  }
\end{table}

 \begin{table}[htbp]
 \centering
\begin{tabular}{ c c |c c c c c }
\hline\hline
&Global &$h_{24}$&$h_{35}$&$h_4^\prime$&$h_{5}^\prime$&$a$
\\
\hline
$h_{24}$&0.983&1.000&-0.493&-0.697&0.135&0.677
\\
$h_{35}$&0.889&-0.493&1.000&-0.165&-0.546&-0.761
\\
$h_4^\prime$&0.971&-0.697&-0.165&1.000&0.501&-0.001
\\
$h_{5}^\prime$&0.954&0.135&-0.546&0.501&1.000&0.780
\\
$a$&0.981&0.677&-0.761&-0.001&0.780&1.000
\\
\hline\hline
\end{tabular}
\caption{ The correlation coefficients between different parameters result from the MINUIT program.  \label{corcoe}}
\end{table}
\vspace{0.5cm}

Most of the data from the present lattice simulations are  for the single-channel cases, 
e.g. $(S,I)=(-1,0), (-1,1), (2,1/2), (0,3/2)$ and 
the $D_s\pi$ channel in the $(S,I)=(1,1)$ case~\cite{Liu:2012zya}. The lattice simulations for 
the $(S,I)=(1,0)$ case, where the $\ds$ appears, are still quite limited. 
Only two data points are given in Refs.~\cite{Mohler:2012na,Mohler:2013rwa,Lang:2014yfa}: 
one is determined with a $N_f=2$ simulation and the other is obtained with $N_f=2+1$. To evaluate the
effect of these   two data points from the $(S,I)=(1,0)$ simulations in the fit, we perform 
two different types of fits. 
The first one is a five-channel fit, denoted as Fit-5C for short, where the data for the 
coupled $DK(I=0)$ channel are not included in the fit, e.g only the data points in the 
$(S,I)=(-1,0), (-1,1), (2,1/2), (0,3/2), (1,1)$ cases from  Ref.~\cite{Liu:2012zya} are considered. 
Compared to the five-channel fit shown in Table~V from Ref.~\cite{Liu:2012zya}, the coupled-channel 
effect for $D_s\pi$ is included here. 
To be more specific, we include the lattice QCD data corresponding to pion masses, $301$~MeV, $364$~MeV 
and $511$~MeV from Ref.~\cite{Liu:2012zya} in the Fit-5C case. The points for $M_\pi>600$~MeV are 
excluded in the fits, since such a large value may challenge a reliable chiral extrapolation. 
The fit results for the parameters are given in the column labeled as Fit-5C in Table~\ref{tab:LECs} 
and  the reproduction of the lattice simulation data can be seen in Fig.~\ref{fig:5CC}. In the 
last panel of this figure, we make a prediction for the $DK$ scattering 
length in the $(S,I)=(1,0)$ case, where in order to make a clear comparison of the lattice 
data from Ref.~\cite{Lang:2014yfa} 
we have used the masses of the $D$ and $D_s$ from that reference to make the plots,  e.g. 
the results in the Ensemble~(2) column in Table~\ref{tab:chirallimit}.  
{Notice that we do not explicitly show the error bands of the predictions in the last panel 
of Fig.~\ref{fig:5CC}, since the error bands are so huge that 
they almost cover the whole region of that panel. } 
It is quite clear that our central predictions prefer the $N_f=2+1$ simulation result over 
the $N_f=2$ one.  This inspires us to perform another type of fit, where we explicitly 
include the $N_f=2+1$ datum from Refs.~\cite{Mohler:2013rwa,Lang:2014yfa} in the fit. 
This fit will be named as Fit-6C in later discussions, as in this case we have six channels in the fit. 
The fit results of the LECs and the subtraction constant are compiled in the column labeled as Fit-6C 
in Table~\ref{tab:LECs} and the comparisons between our chiral extrapolations using 
Eq.~(\ref{eq:chiralextra}) and lattice QCD data are displayed in Fig.~\ref{fig:6CC}. 
Due to the inclusion of the coupled channel effects, improvements are achieved in the 
sense that the error bars become smaller, especially for the results in Fit-6C. Nevertheless, 
our results and those in Ref.~\cite{Liu:2012zya} are compatible within the uncertainties and 
the resulting values for LECs are quite similar, see Table~\ref{tab:LECs} for  comparisons.  
The corresponding descriptions of the lattice QCD data are all of good quality with a $\chi^2$ per 
d.o.f  around 1.0,  which can also be seen from the plots in Fig.~\ref{fig:5CC} and Fig.~\ref{fig:6CC}. 
In Figs.~\ref{fig:5CC} and \ref{fig:6CC}, the shaded bands represent the variation of the scattering 
lengths with the LECs whose values vary within a 1-$\sigma$ uncertainty.

 \begin{figure}[ht]
   \centering
   \includegraphics[width=1\textwidth]{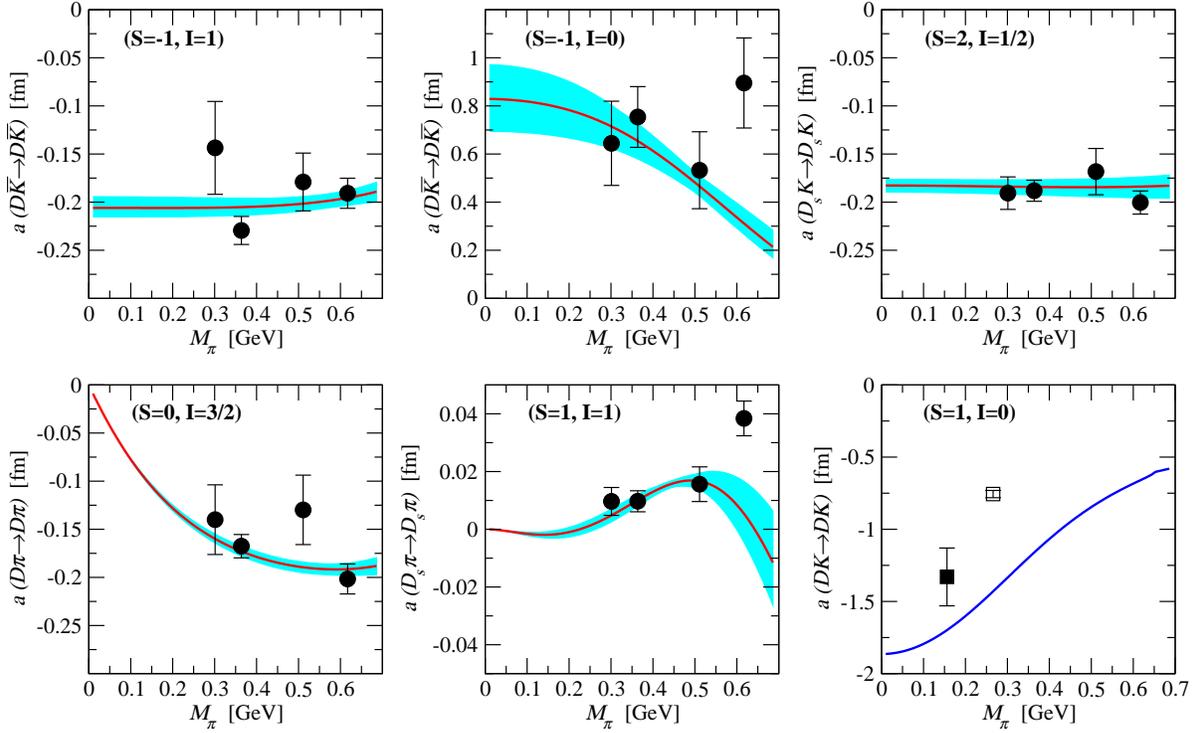} 
  \caption{(Color online) 5-channel Fit (Fit-5C). The bands represent the variation of the 
scattering lengths with the LECs varying within 1-$\sigma$ uncertainty. The black solid 
circles with error bars stand for the lattice QCD data from Ref.~\cite{Liu:2012zya}. The open 
and solid squares with error bars denote the $N_f=2$ and $N_f=2+1$ lattice QCD data from 
Ref.~\cite{Mohler:2013rwa}, respectively. The blue solid line in the ($S=1,I=0$) panel is 
our prediction, see the text for details. }
   \label{fig:5CC}
\end{figure}

\begin{figure}[ht]
   \centering
   \includegraphics[width=1\textwidth]{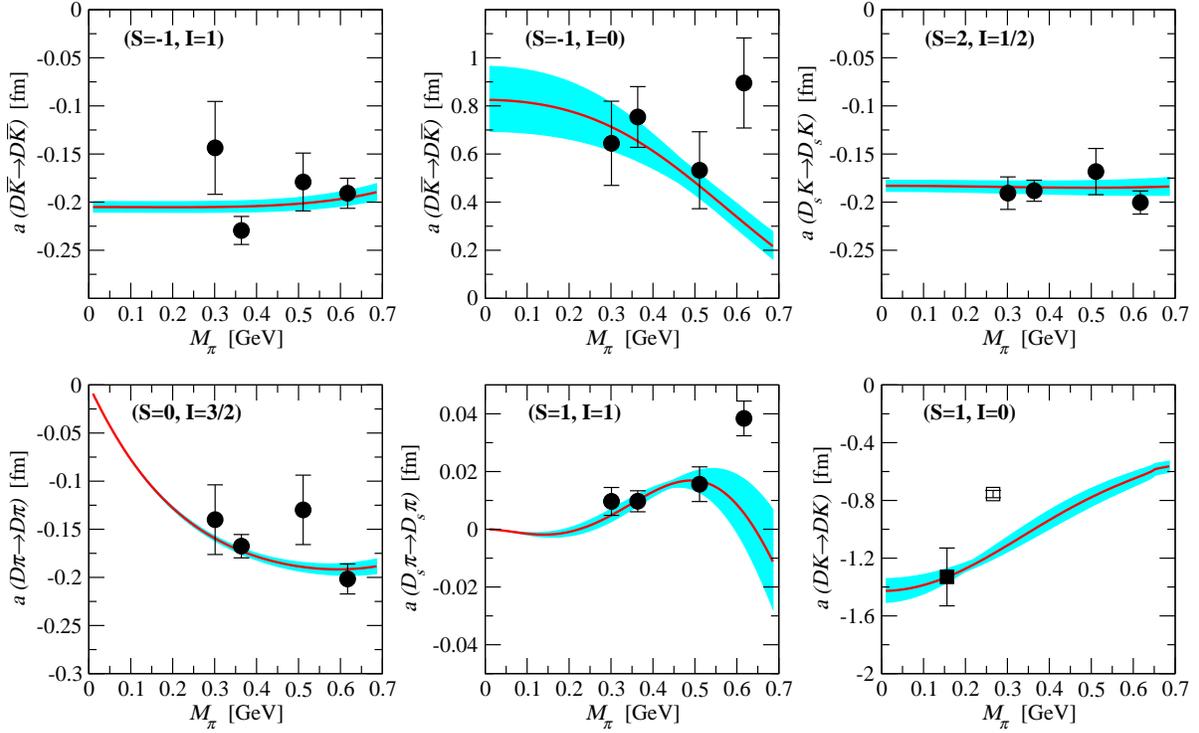} 
  \caption{(Color online) 6-channel fit (Fit-6C). The bands represent the variation of 
the scattering lengths with the LECs varying within 1-$\sigma$ uncertainty. 
  See Fig.~\ref{fig:5CC} for further details. }
   \label{fig:6CC}
\end{figure}

\begin{figure}[htbp]
   \centering
   \includegraphics[width=0.8\textwidth]{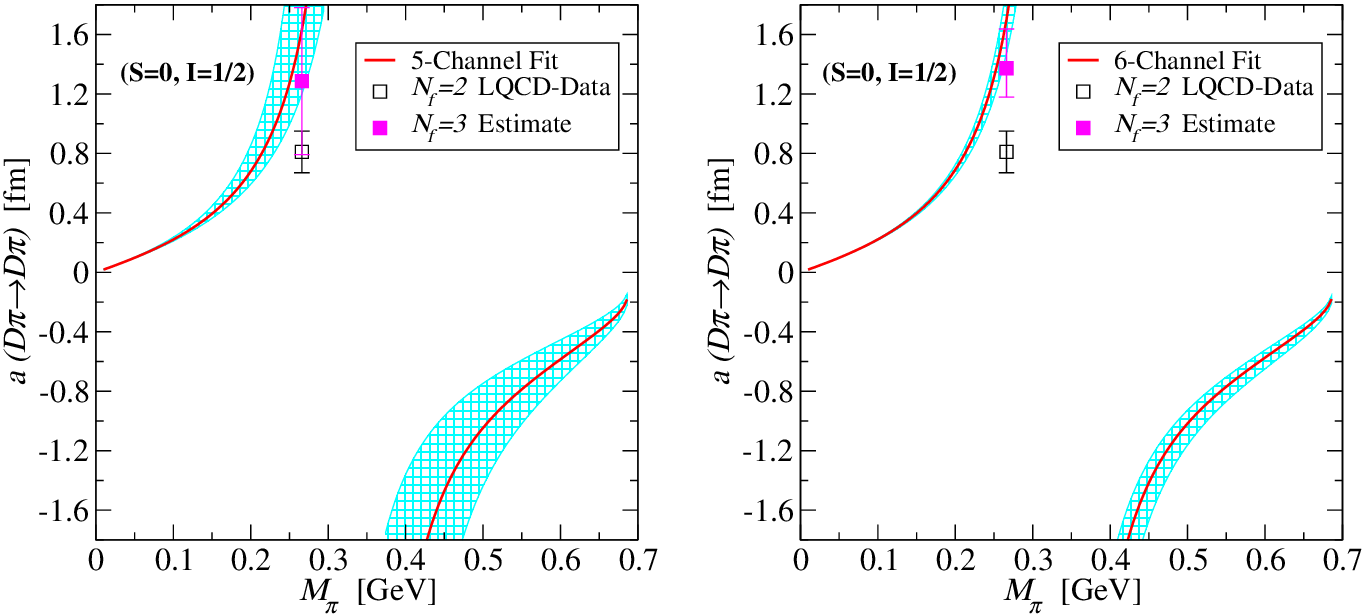} 
  \caption{(Color online) Predictions for the $D\pi\to D\pi$ scattering lengths in the $(S=0,~I=1/2)$ 
channel. The red lines with the cyan grid bands represent the chiral extrapolations of the scattering 
lengths with the masses specified by the last column in Table~\ref{tab:chirallimit} and the LECs 
varying within their 1-$\sigma$ uncertainties shown in Table~\ref{tab:LECs}. The open black square with 
error bar denotes the $N_f=2$ lattice QCD data taken from Ref.~\cite{Mohler:2012na}. The magenta 
solid square with error bar is our estimate based on the masses of Ensemble~(2) in 
Table~\ref{tab:chirallimit} and LECs in Table~\ref{tab:LECs}.}
   \label{fig:Dpi}  
\end{figure}

Our prediction for the chiral extrapolation of $D\pi$ scattering length with $(S,I)$=$(0,1/2)$ 
is given in Fig.~\ref{fig:Dpi},  where we have used the masses in the column labeled by 'Physical' 
in Table~\ref{tab:chirallimit}. In addition, with the LECs obtained in Table~\ref{tab:LECs} and 
masses specified at their corresponding physical values, we predict all the physical $S$-wave 
scattering lengths in Table~\ref{tab:SL}.

 \begin{table}[htbp]
 \centering
{\small 
\begin{tabular}{ c c | c c }
\hline\hline
$(S,\, I)$&Channels& Fit-5C & Fit-6C
\\ \hline
$(-1,0)$& $D\bar{K}\to D\bar{K}$&$1.27^{+0.49}_{-0.36}$&$1.26^{+0.46}_{-0.32}$
\\
$(-1,1)$&$D\bar{K}\to D\bar{K}$&$-0.21^{+0.02}_{-0.01}$&$-0.21^{+0.01}_{-0.01}$
\\
$(2,\frac{1}{2})$&$D_sK\to D_s K$&$-0.19^{+0.01}_{-0.01}$& $-0.19^{+0.01}_{-0.01}$
\\
$(0,\frac{3}{2})$&$D\pi\to D\pi$&$-0.101^{+0.003}_{-0.003}$&$-0.101^{+0.001}_{-0.001}$
\\
$(1,1)$&$D_s\pi\to D_s\pi$&$0.004^{+0.001}_{-0.001}$&$0.004^{+0.001}_{-0.001}$
\\
&$DK\to DK$&$0.06^{+0.03}_{-0.03}+i\, 0.17^{+0.02}_{-0.01}$&$0.06^{+0.03}_{-0.03}+i\,0.17^{+0.01}_{-0.01}$
\\
(1,0)&$DK\to DK$&$-0.92^{+0.22}_{-0.40}$&$-0.89^{+0.06}_{-0.10}$
\\
&$D_s\eta\to D_s\eta$&$-0.27^{+0.01}_{-0.01}+i\,0.03^{+0.01}_{-0.01}$&$-0.27^{+0.01}_{-0.01}+i\,0.03^{+0.01}_{-0.01}$
\\
&$D_s\eta^\prime\to D_s\eta^\prime$&$-0.22^{+0.03}_{-0.01}+i\,0.01^{+0.01}_{-0.01}$&$-0.22^{+0.01}_{-0.01}+i\,0.01^{+0.01}_{-0.01}$
\\
$(0,\frac{1}{2})$&$D\pi\to D\pi$&$0.35^{+0.04}_{-0.02}$&$0.35^{+0.01}_{-0.01}$
\\
&$D\eta\to D\eta$&$0.02^{+0.06}_{-0.04}+i\,0.03^{+0.03}_{-0.01}$&$0.02^{+0.02}_{-0.02}+i\,0.03^{+0.01}_{-0.01}$
\\
&$D_s\bar{K}\to D_s\bar{K}$&$-0.05^{+0.04}_{-0.06}+i\,0.35^{+0.07}_{-0.03}$&$-0.05^{+0.02}_{-0.02}+i\,0.35^{+0.04}_{-0.03}$
\\
&$D\eta^\prime\to D\eta^\prime$&$0.16^{+0.64}_{-0.22}+i\,0.05^{+0.26}_{-0.03}$&$0.34^{+0.31}_{-0.14}+i\,0.04^{+0.12}_{-0.02}$
\\
\hline\hline
\end{tabular}
\caption{\label{tab:SL} Predictions of the scattering lengths using the parameters from 
Table~\ref{tab:LECs} together with the physical masses for the charmed and light pseudoscalar mesons.}
}
\end{table}

\section{Pole analyses on the charm scalar states}\label{sect.pole}

\subsection{ Pole contents in the physical situation}
Experimentally observed bound states and resonances correspond to the poles of the 
partial-wave amplitudes in physical and unphysical Riemann sheets~(RS),  respectively. 
In this section, we utilize the previously determined LECs and subtraction constant to 
study the pole contents in the charmed mesons and pNGBs scattering amplitudes. 
In our formalism, different Riemann sheets are characterized by the sign of imaginary part 
of the loop function $g(s)$ in Eq.~\eqref{gfunc}. 
Each $g(s)$  is associated with two sheets. By default the expression in Eq.~\eqref{gfunc} defines 
the first/physical Riemann sheet.  To reverse the sign of the imaginary part of the $g(s)$ function 
in Eq.~\eqref{gfunc}, one then analytically extrapolates to the unphysical Riemann sheet. 
For scattering processes in question, different Riemann sheets can be accessed by choosing the 
proper signs of the imaginary parts of the $g_i(s)$ functions, 
with $g_i(s)$ calculated from the two intermediate states in the $i^{\rm th}$-channel. The first 
Riemann sheet will be labeled as 
$(+,+,+,...)$.  The second and third sheets can be accessed by reversing the signs of the 
imaginary parts of the $g(s)$ functions defined at the first and second thresholds, which will 
be denoted by  $(-,+,+,...)$ and $(-,-,+,...)$, in order. 
Besides the pole positions, the residues for a given pole calculated in the partial-wave 
amplitudes in different channels can also provide useful information, since the residues 
correspond to the coupling strengths between the pole and its interacting $D\phi$ modes. 

Since the results in Table~\ref{tab:LECs} for Fit-5C and Fit-6C are quite similar, we only 
present the pole analyses for the Fit-6C case in the following. Nevertheless, we stress that 
we have explicitly verified that the poles and their residues from the two fit results look 
indeed very similar.  The pole positions and their residues in different channels, which are 
calculated by using the physical masses for the charmed mesons and pNGBs, are summarized 
in Table~\ref{tab:pole6c}. Some remarks about the comparisons of our results for the charm scalar meson spectra with 
the previous ones in literature and experimental measurements are in order. 


{A virtual-state pole at around 2.3~GeV is found in the single channel with $(S,I)=(-1,0)$. 
In the  $(S,I)=(0,3/2)$ channel a very broad resonance with the width around $500$~MeV is obtained. 
Such a broad resonance is difficult to be verified in experiments. }

For the $(S,I)=(1,1)$ case with two coupled channels, we find one broad 2nd-Riemann-sheet pole 
above the $DK$ threshold and one  broad 3rd-Riemann-sheet pole below the $DK$ threshold. The 
former one is mainly coupled to $DK$ and the latter is coupled more or less equally to both 
$D_s\pi$ and $DK$.  Our results are different from those in Ref.~\cite{Wang:2012bu}, especially 
that our width for the 2nd-Riemann-sheet pole is much larger than that in the previous reference. 
{These differences are presumably  caused by the fact that the results in Ref.~\cite{Wang:2012bu} are obtained replying on the preliminary scattering length data from Ref~\cite{Liu:2008rza}, which somewhat differ from the final published data that we used in this work,  given in Ref.~\cite{Liu:2012zya}.
}
It is quite interesting to  point out that an enhancement in the $D^0K^+$ invariant mass 
around $2350$-$2500$~MeV was very recently observed~\cite{Purohit:2015zka}, which may 
be an evidence of the existence of the pole with $\sqrt{s_{\rm pole}}=2466_{-27}^{+32}-i\,271_{-5}^{+4}$ 
in Table~\ref{tab:pole6c}. 

In the $(S,I)=(1,0)$ case, a bound-state pole at $2321_{-3}^{+6}$ is found and it corresponds to 
the physical $D_{s0}^\ast(2317)$ state. We further verify that the $\ds$ is most strongly coupled 
to the $DK$ system, as can be seen by its residues in Table~\ref{tab:pole6c}. In addition, we 
also find a virtual-state pole just below the $DK$ threshold on the second Riemann sheet. It is 
interesting to stress that the virtual pole is even closer to the $DK$ threshold than the 
bound $\ds$ state. 

In the $(S,I)=(0,1/2)$ channel, there are two poles: $\sqrt{s_{\rm pole}}=2114_{-3}^{+3}-i\,111_{-7}^{+8}$ 
and $\sqrt{s_{\rm pole}}=2473_{-22}^{+29}-i\,140_{-7}^{+8}$, located on the second and third Riemann sheets, 
respectively. The second-Riemann-sheet pole is consistent with the observations made in by 
Refs.~\cite{Guo:2006fu,Wang:2012bu,Lutz:2003ac, Hofmann:2003je}. This pole can be regarded as a 
broad $D_0^\ast~(0,\frac{1}{2})$ state strongly coupled to the $D\pi$ threshold.  On the other hand, 
the third-Riemann-sheet pole, which is most strongly coupled to $D_s\bar{K}$, has a much broader 
width compared to the previous results~\cite{Wang:2012bu,Lutz:2003ac,Hofmann:2003je} and our 
determination of its mass and width is close to those of $D_0^\ast(2460)$ announced by PDG~\cite{pdg14}. 


 \begin{table}[htbp]
 \centering
\begin{tabular}{ c c | c c c c }
\hline\hline
$(S,\, I)$&RS& $\sqrt{s_{pole}}$~~[MeV] & ${\rm |Residue|}^{1/2}$~[GeV]&\multicolumn{2}{c}{Ratios}
\\ \hline
$(-1,0)$&II&$2333_{-36}^{+15}$&$7.45_{-1.38}^{+3.56}(D\bar{K})$&&
\\
$(0,\frac{3}{2})$&II&$2033_{-3}^{+3}-i\,251_{-3}^{+3}$&$6.64_{-0.04}^{+0.04}(D\pi)$&&
\\
$(1,1)$&II& $2466_{-27}^{+32}-i\,271_{-5}^{+4}$&$6.95_{-0.37}^{+0.60}(D_s\pi)$&$1.72_{-0.15}^{+0.12}(DK/D_s\pi)$&
\\
&III&$2225_{-9}^{+12}-i\,178_{-17}^{+19}$&$7.35_{-0.13}^{+0.19}(D_s\pi)$&$0.80_{-0.04}^{+0.04}(DK/D_s\pi)$&
\\
(1,0)&I&$2321_{-3}^{+6}$&$9.30_{-0.12}^{+0.04}(DK)$&$0.77_{-0.02}^{+0.02}(D_s\eta/DK)$& $0.43_{-0.13}^{+0.15}(D_s\eta^\prime/DK)$
\\
&II&$2356_{-1}^{+1}$&$2.85_{-0.13}^{+0.08}(DK)$&$0.69_{-0.01}^{+0.01}(D_s\eta/DK)$&$0.38_{-0.11}^{+0.12}(D_s\eta^\prime/DK)$
\\
$(0,\frac{1}{2})$&II& $2114_{-3}^{+3}-i\,111_{-7}^{+8}$&$9.66_{-0.13}^{+0.15}(D\pi)$&$0.31_{-0.03}^{+0.03}(D\eta/D\pi)$&$0.46_{-0.02}^{+0.02}(D_s\bar{K}/D\pi)$
\\
&&&&$0.49_{-0.08}^{+0.08}(D\eta^\prime/D\pi)$&
\\
&III&$2473_{-22}^{+29}-i\,140_{-7}^{+8}$&$5.36_{-0.28}^{+0.40}(D\pi)$&$1.09_{-0.05}^{+0.06}(D\eta/D\pi)$&$2.12_{-0.08}^{+0.06}(D_s\bar{K}/D\pi)$
\\
&&&&$1.12_{-0.16}^{+0.18}(D\eta^\prime/D\pi)$&
\\
\hline\hline
\end{tabular}
\caption{\label{tab:pole6c} Poles and their residues based on Fit-6C in Table~\ref{tab:LECs}. 
Physical masses for the charmed and light pseudoscalar mesons are used to obtain the results 
in this table. The Riemann sheets on which the poles are located are indicated in the second column. 
In the last column, we give the ratios of the residues with respect to the first threshold. }
\end{table}

\subsection{Pole trajectories with varying pion mass}

Due to the limited experimental measurements of the charm scalar mesons, lattice simulations 
provide a possible way to verify the poles from our analyses. 
Therefore it is interesting to further probe the pole trajectories with varying pion mass $M_\pi$, 
which can be useful for comparisons with  future lattice results. From the theoretical point of view, 
the $M_\pi$ trajectories of the various poles can offer us further insights into the properties of 
different hadron states, as discussed e.g. in Ref.~\cite{Hanhart:2014ssa}. 
Before investigating the $M_\pi$ trajectories of the various poles, we first show how the 
masses of the $K, \eta, \eta', D, D_s$, their thresholds and the mixing angle $\theta$ change with  
varying pion mass in Fig.~\ref{fig:thresholdMpi}. We point out that in the discussions below we 
have used the physical strange quark mass, i.e. the values in the last column in 
Table~\ref{tab:chirallimit}. All of the masses considered here are increased when enlarging 
the pion masses, and the mixing angle $\theta$ tends to increase as well.  Around $M_\pi \sim 700$~MeV, 
it is clear from Fig.~\ref{fig:thresholdMpi} that the masses of $D$ and $D_s$ become equal and the 
mixing angel $\theta$ approaches zero. 
This phenomenon is not a surprise, since when $M_\pi \simeq 690$~MeV, the pion and kaon masses turn out 
to be degenerate, indicating the exact $SU(3)$-flavor symmetry in that region.

The $M_\pi$ trajectories for the $\ds$ meson can be seen in Fig.~\ref{fig:mpiDs0pole}.
In the left panel, together with the varying thresholds $M_D+M_K$, we give the results for 
the pole positions of $\ds$ on the first Riemann sheet, 
which are identified as its masses. In the right panel, we show the binding energies 
which are the gaps between the thresholds and the pole positions. 
The red solid lines in Fig.~\ref{fig:mpiDs0pole} stand for our final results of the $U(3)$ chiral 
theory with realistic descriptions of $\eta$ and $\eta'$, while the blue dashed lines labeled 
as $SU(3)$ are calculated by approximating our final expressions to mimic the $SU(3)$ case, 
which are obtained by setting  $M_0 \to \infty$~\footnote{Strictly speaking, one would have to readjust 
the LECs, too.}. In this limit, the heavy singlet $\eta_0$ and the octet $\eta_8$ will be 
decoupled, and one can identify the octet $\eta_8$ as the physical $\eta$, as done in 
$SU(3)$ $\chi$PT~\cite{Gasser:1984gg}. In addition, when $M_0 \to \infty$ the effects from the 
channels with the heavy singlet $\eta_0$ will be much suppressed due to their far distance form the 
thresholds. The small variances between the $U(3)$ and $SU(3)$ lines in Fig.~\ref{fig:mpiDs0pole} 
indicate that the final results for $\ds$ are not very sensitive to the treatment of the $\eta$ 
and the $\eta'$ at least for low pion masses. This also explicitly justifies the previous study 
of $\ds$ in 
Refs.~\cite{Lutz:2003ac,Hofmann:2003je,Guo:2006fu,Guo:2009ct,Cleven:2010aw,Wang:2012bu,Geng:2013vwa} 
based on the $SU(3)$ treatment of $\eta$.

The most important conclusion obtained from Fig.~\ref{fig:mpiDs0pole} is that the
$\ds$ always stays a bound state below the $DK$ threshold for a wide range of $M_\pi$. 
In contrast, the $M_\pi$ trajectories of the pole around $2.1$~GeV on the second 
Riemann sheet with $(S,I)=(0,1/2)$ in Table~\ref{tab:pole6c} are found to be quite  
complicated, as shown in Fig.~\ref{fig:mpis0i12lpole}. By increasing the value of 
$M_\pi$, we first see that both the real and imaginary parts of this pole 
tend to decrease on the second Riemann sheet. At some point around $M_\pi\sim 
2 M_\pi^{\rm phys}$, the real part of this pole becomes lower than the threshold of 
$D\pi$, but its imaginary part is still nonzero. When $M_\pi$ equals to 288~MeV, 
the pair of resonance poles meets at the point below the $D\pi$ threshold on 
the real axis and becomes two virtual states on the second Riemann sheet. If we 
further increase the value of the pion mass, one of the virtual poles goes further 
away from the threshold towards to the left side on the real axis, while the other 
one moves closer to the threshold on the right side and it becomes a bound state 
on the first Riemann sheet 
when $M_\pi >$336~MeV. If we keep increasing the value of $M_\pi$, both the 
virtual and the bound state move further away from the threshold to the left side. 
It is very interesting to point out that the behavior for the broad charm scalar 
pole around 2.1~GeV with $(S,I)=(0,1/2)$ looks quite similar to the one of 
the $\sigma$ resonance $f_0(500)$ discussed 
in Refs.~\cite{Hanhart:2008mx,Hanhart:2014ssa}. 

\begin{figure}[ht]
\begin{center}
   \includegraphics[width=0.9\textwidth]{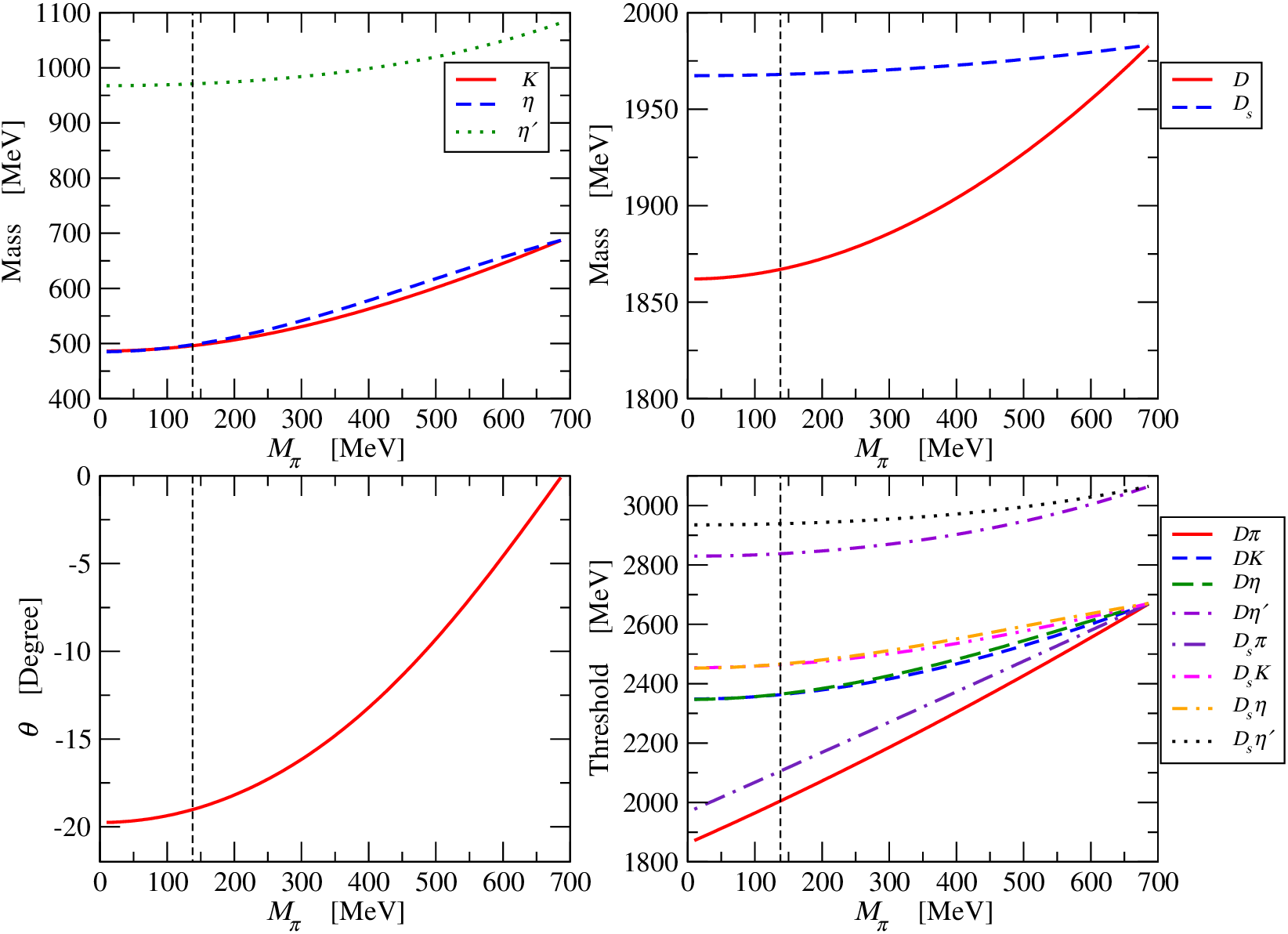} 
  \caption{Masses for the pseudo-Goldstones bosons and the charmed D mesons, 
    the LO $\eta$-$\eta^\prime$ mixing angle and the thresholds as functions of $M_\pi$.}
   \label{fig:thresholdMpi}
    \end{center}
\end{figure}

The pole trajectories with varying $M_\pi$ for the charm scalar pole 
around $2.4$~GeV in the third Riemann sheet with $(S,I)=(0,1/2)$ are 
shown in Fig.~\ref{fig:mpis0i12hpole}. Compared to the trajectories in 
Fig.~\ref{fig:mpis0i12lpole}, the dependences of the pole around $2.4$~GeV 
with $M_\pi$ are much weaker and only mild changes of the pole positions are 
observed over a wide range of pion masses from 100 to 700~MeV.

\begin{figure}[ht]
\begin{center}
   \includegraphics[width=0.9\textwidth]{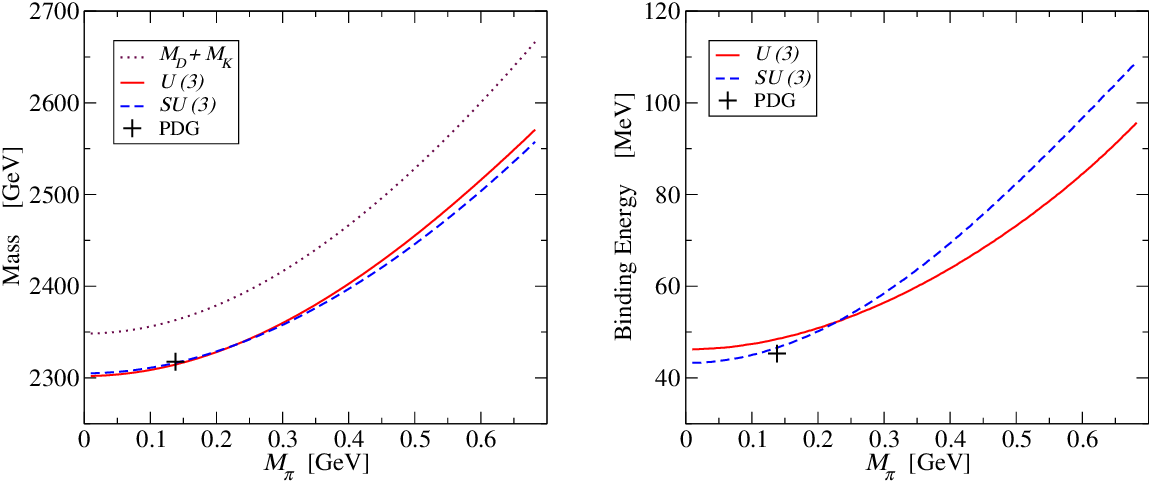} 
  \caption{Properties of $D_{s0}(2317)$ pole masses when varying 
           $M_\pi$: $U(3)$ v.s. $SU(3)$. }
   \label{fig:mpiDs0pole}
   \end{center}
\end{figure}

\begin{figure}[ht]
\begin{center}
   \includegraphics[width=0.9\textwidth]{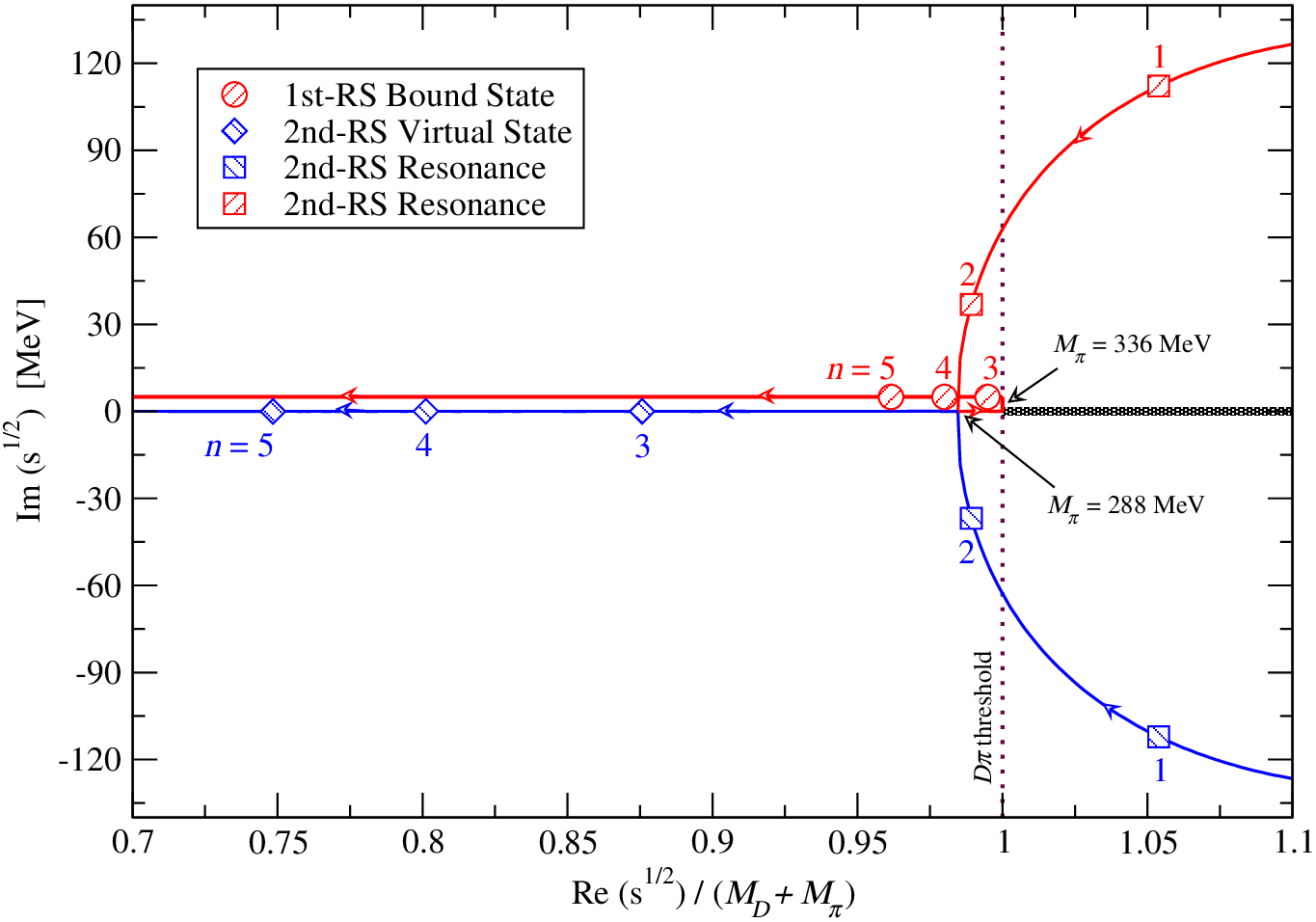} 
  \caption{Trajectories of the $(S=0,I=1/2)$ resonance at around 2.1~GeV 
           with varying $M_\pi$. $n$ is defined by $M_\pi=n\,M_\pi^{\rm phys}$}
   \label{fig:mpis0i12lpole}
   \end{center}
\end{figure}

\begin{figure}[ht]
\begin{center}
   \includegraphics[width=0.9\textwidth]{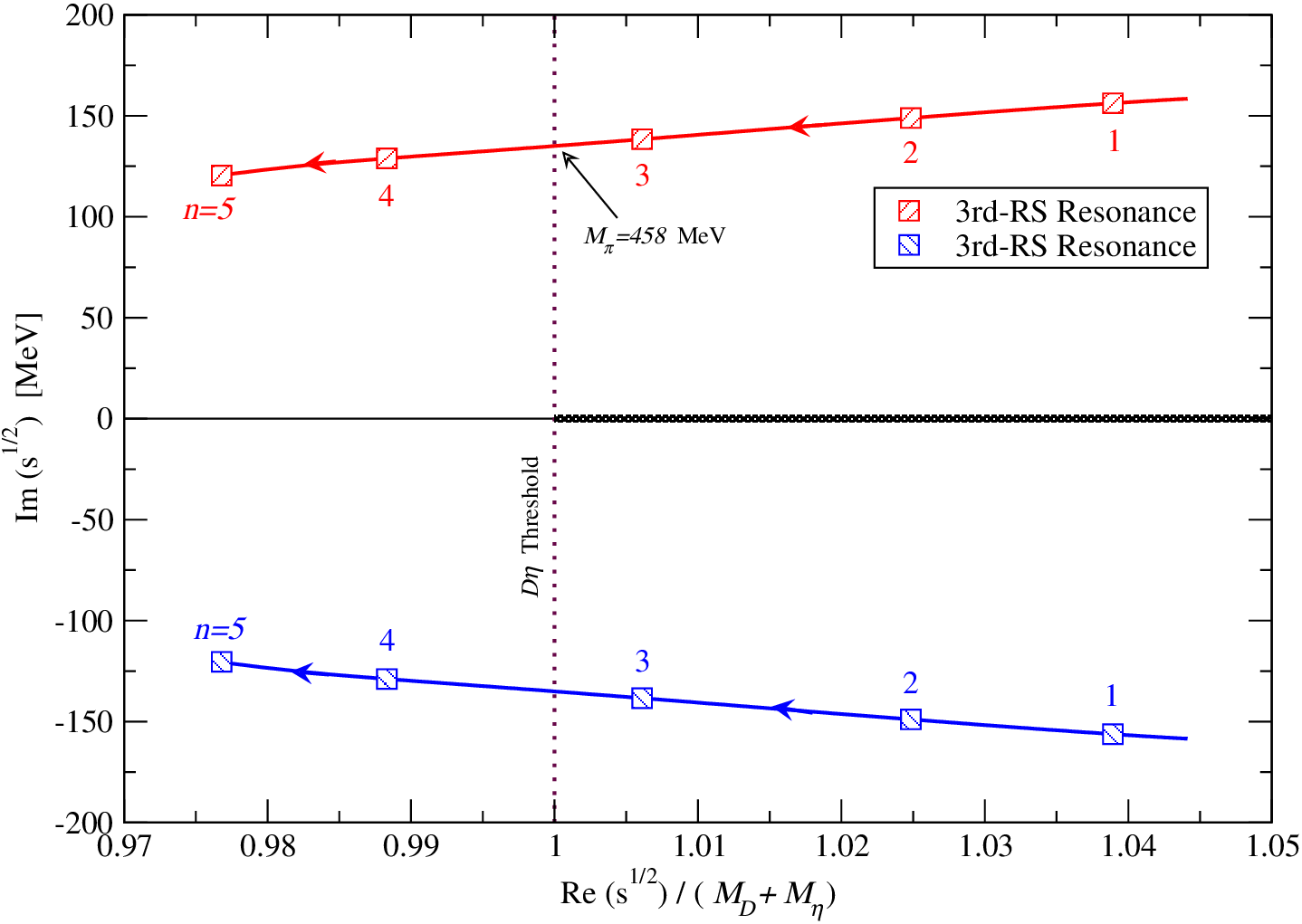} 
  \caption{Trajectories of the $(S=0,I=1/2)$ resonance at around 2.4~GeV 
           with varying $M_\pi$. $n$ is defined by $M_\pi=n\,M_\pi^{\rm phys}$}
   \label{fig:mpis0i12hpole}
   \end{center}
\end{figure}

\subsection{$N_C$ trajectories for the charm scalar mesons}

The $N_C$ trajectory of a resonance/bound-state pole can provide us useful 
information about the inner structure of the particle. 
Extensive studies on the $N_C$ trajectories for light-flavor resonances, 
such as $f_0(500)$, $f_0(980)$, $\rho(770)$, $a_1(1260)$ etc, have been 
carried out in 
literature~\cite{Sun:2005uk,Pelaez:2006nj,Dai:2011bs,Guo:2011pa,Guo:2012yt,Pelaez:2009eu,Nagahiro:2011jn}. 
As  discussions on the $N_C$ properties of the heavy-light mesons are still rare, 
one of the key goals of this work is to fill this gap. 
We point out that in order to make a reliable study of the $N_C$ trajectories 
for a given state, not only the $N_C$ scaling of the couplings 
of the scattering amplitudes, but also the $N_C$ running of the masses of the 
intermediate particles involved in the scattering, need to be carefully considered. 
As mentioned in the Introduction,  the singlet $\eta_0$, with its large mass due 
to the QCD $U(1)_A$ anomaly effect, 
should be properly taken into account when discussing the $N_C$ dependences, 
as the $U(1)_A$ anomaly  is $1/N_C$ suppressed when $N_C \to \infty$ and $\eta_0$ 
then becomes the ninth pNGB in the chiral and large $N_C$ limits.

For the $N_C$ behaviors of the LECs in Eq.~(\ref{nlolag}), their leading-order 
$N_C$ scaling can be obtained by counting the number of 
traces in the corresponding operators. We take $h_0$ and $h_1$ as examples to 
illustrate this. The operator accompanied by $h_1$ has the same number 
of traces as the mass operator in Eq.~(\ref{lolag}), therefore $h_1$ has the 
same $N_C$ scaling as the bare mass squared $\overline{M}_D^2$. 
The leading scaling of $h_0$ is one more power of $1/N_C$ suppressed since 
there is one additional trace in this operator.  
From the large $N_C$ point of view, the mass of any $\bar{q}q$ meson behaves as a 
constant~\cite{Nc1,Nc2} and, 
therefore, the 
leading $N_C$ scaling for $\overline{M}_D$, $M_\pi$ and $M_K$ is $\cO(1)$. As a 
result, the leading $N_C$ scaling of $h_1$ is also $\cO(1)$, while 
$h_0$ should be counted as $\cO(1/N_C)$. Similar rules can be applied to other 
operators in Eq.~(\ref{nlolag}) and we summarize 
their leading $N_C$ scaling as follows{\yao  }
\begin{equation}\label{Eq:hisNc}
 h_1, h_3, h_5 \sim \cO(1), \quad h_0, h_2, h_4 \sim \cO(1/N_C)\,. 
\end{equation}
The $N_C$ scaling of the pNGB decay constant $F$ in the chiral limit is 
$\cO(\sqrt{N_C})$~\cite{Gasser:1984gg,Manohar:1998xv} and the 
singlet $\eta_0$ mass squared $M_0^2$ is counted as 
$\cO(1/N_C)$~\cite{ua1nc0,ua1nc1,ua1nc2,Manohar:1998xv}. For the subtraction constant $a$ 
in Eq.~\eqref{gfunc}, 
a reasonable assignment for its leading $N_C$ scaling is $\cO(1)$ as 
argued in Ref.~\cite{Guo:2012yt} and we will also use this.
With these assignments of the $N_C$ scaling for different parameters, 
we can calculate the $N_C$ running for other quantities.

In Fig.~\ref{fig:thresholdNc}, we  show the $N_C$ scaling of the masses of 
the pNGBs and charmed $D$ mesons, the $\eta$-$\eta'$ mixing angle and the 
relevant thresholds. A striking phenomenon from the first panel in 
Fig.~\ref{fig:thresholdNc} is that the $\eta$ mass significantly decreases 
when $N_C$ is increased and it tends to be equal to the pion mass in the 
large $N_C$ limit. Contrary to this, when increasing the values of $N_C$, the mass 
of the $\eta'$ decreases  from the beginning, but it still has a relatively 
large mass around 700~MeV in the large $N_C$ limit.  These behaviors can be 
analytically understood 
from the leading order mixing formulas in Eqs.~\eqref{defmetab2}, \eqref{defmetaPb2} and \eqref{deftheta0}. In the large $N_C$ limit, one has $M_0 \to 0$ and 
in this limit one can easily obtain from Eqs.~\eqref{defmetab2}, \eqref{defmetaPb2} and \eqref{deftheta0} that 
\begin{equation}
 M_\eta= M_\pi, \quad  M_{\eta'} = \sqrt{2M_K^2 - M_\pi^2}, \quad  \arcsin\theta = -\sqrt{\frac{2}{3}} \,,
\end{equation}
which perfectly explains the results in the two panels on the left side of 
Fig.~\ref{fig:thresholdNc}. The results are consistent with the findings 
in Ref.~\cite{Weinberg:1975ui}.  
The masses of the charmed $D_{(s)}$ mesons depend on the two LECs $h_0$ and $h_1$, 
as explicitly given in Eqs.~\eqref{mdmpi} and \eqref{mdms}. 
The final results for the $N_C$ running of $M_D$ and $M_{D_s}$ are given in the 
top right panel in Fig.~\ref{fig:thresholdNc}, 
which basically behave like constants when varying $N_C$. As a result, the 
$N_C$ running of the thresholds as given in the bottom right panel 
is mainly caused by the running of $M_\eta$ and $M_{\eta'}$. An important finding 
is that the order of the thresholds $M_{D_s} + M_\eta$ and $M_D+M_K$ is  
reversed and the former becomes lower than the latter for $N_C>7$.

The pole trajectories of the $\ds$ with varying $N_C$ are given {in the left panel of}  
Fig.~\ref{fig:NcDs0pole} for the leading $N_C$ scaling of $h_i$. 
In order to highlight the relative position of the pole and the $DK$ threshold, 
we have normalized the units of the real axis in terms of $M_D+M_K$. 
The physical $\ds$ state is represented by 
the bound-state pole at $N_C=3$ on the first Riemann sheet, i.e. the left most 
point in the real axis in Fig.~\ref{fig:NcDs0pole}. We also 
find a virtual-state pole below the $DK$ threshold on the real axis on 
the second Riemann sheet at $N_C=3$, which is even closer to the threshold 
than the bound-state pole. The explicit values of the pole positions at $N_C=3$ 
can be seen in Table~\ref{tab:pole6c}. When increasing the values of $N_C$, 
we observe that both the bound-state and virtual-state poles approach to threshold, 
and they finally meet at the threshold around $N_C=6$. 
By further increasing the values of $N_C$, the bound-state and virtual-state 
poles move into the complex plane on the second Riemann sheet and become 
a pair of resonance poles. At $N_C=7$ we see a kink structure in the trajectories 
and by increasing $N_C$ afterwards the resonance poles move 
deeper and deeper into the complex plane, with increasing real and imaginary parts. 
The trajectory of $\ds$ for large values of $N_C$ is clearly different from 
the behavior  of a standard quark-antiquark meson in QCD, which would fall 
down to the real axis with the mass behaving as a constant and the width 
scaling as $1/N_C$~\cite{Manohar:1998xv}.  
{ In the left panel of Fig.~\ref{fig:Ncs0i12pole}, we show the $N_C$ trajectories of the poles 
in the $(S,I)=(0,1/2)$ case by considering the leading $N_C$ scaling laws of $h_i$ in Eq.~\eqref{Eq:hisNc}.  }
Compared to the $\ds$ case in Fig.~\ref{fig:NcDs0pole}, 
the trajectories of the two poles in the $(S,I)=(0,1/2)$ channel show a much 
simpler behavior. For large values of $N_C$, both the real and imaginary parts 
of the two poles tend to increase and they keep moving into the complex plane 
instead of falling down to the real axis. This behavior look similar to that 
of the $\ds$ for large values of $N_C$, indicating that the two charm scalar 
resonances with $(S,I)=(0,1/2)$ do not seem to be a standard quark-antiquark 
meson at large $N_C$.

\begin{figure}[ht]
   \centering
   \includegraphics[width=0.9\textwidth]{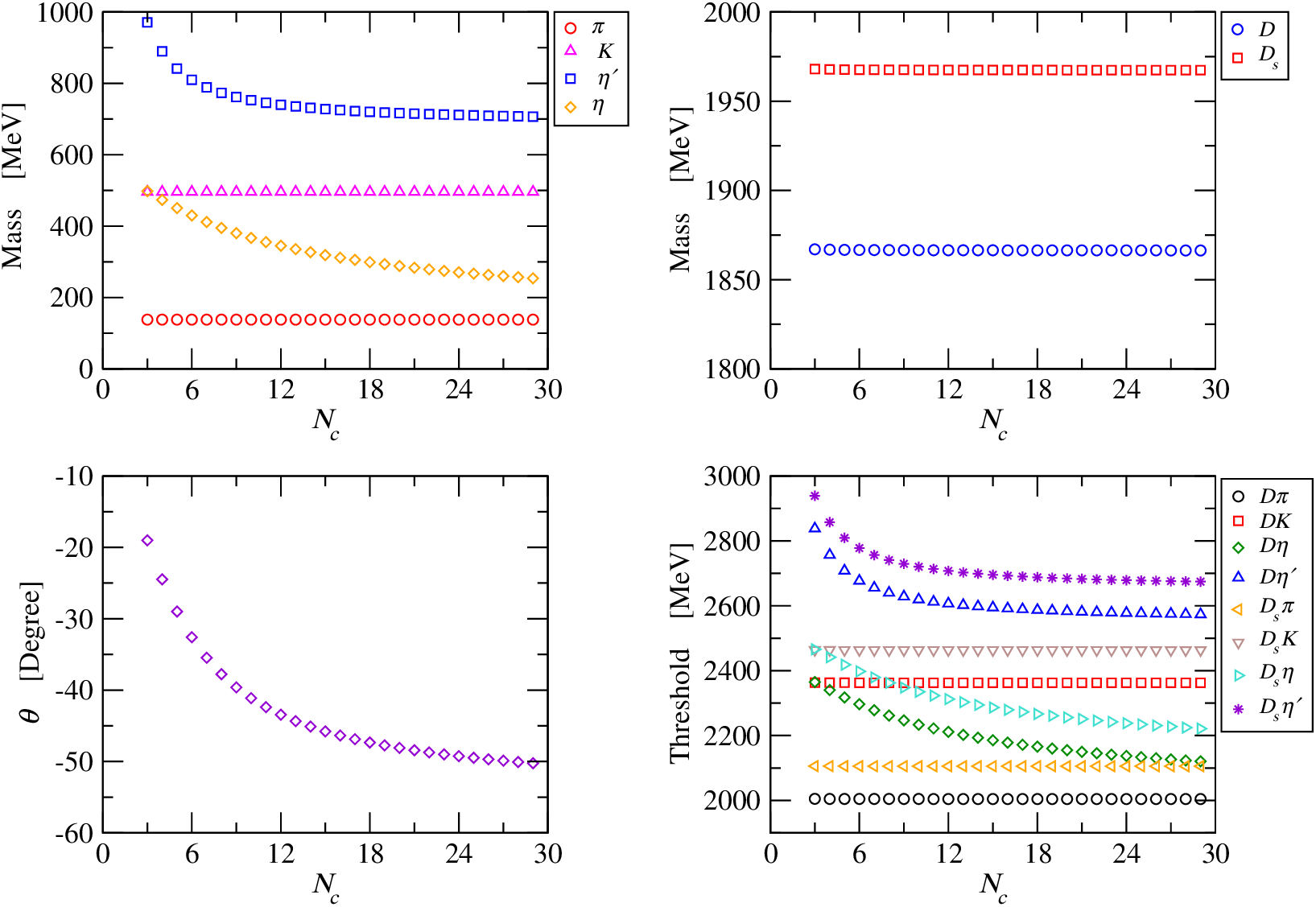} 
  \caption{Masses for the pNGBs and charmed D mesons, the LO $\eta$-$\eta^\prime$ 
           mixing angle and the relevant thresholds as functions of $N_C$ from 
           3 to 30 in steps of one unit.}
   \label{fig:thresholdNc}
\end{figure}

{
The previous $N_C$ discussions are based on using the leading $N_C$ scaling laws of the LECs $h_i$ 
given in Eq.~\eqref{Eq:hisNc}. In order to check the stability of our conclusions on the $N_C$  dependence of the
poles, we follow Ref.~\cite{Guo:2012yt} to include the subleading $N_C$ scaling effects for the 
$h_i$. The explicit formulae read
\bea\label{ncnlo1}
h_{i}(N_c)= 
   {h_i}(N_c=3)\times\left\{1+\frac{h_i(N_c=3)-{h_i}(N_c=\infty)}{h_i(N_c=3)}\left(\frac{3}{N_c}-1\right)\right\}\,, \quad  i=1,3,5
\eea
and 
\bea\label{ncnlo2}
h_{i}(N_c)=\frac{3}{N_c}{h_i}(N_c=3)\times\left\{1+\frac{h_i(N_c=3)-h_i^{Nor}(N_c=\infty)}{h_i(N_c=3)}\left(\frac{3}{N_c}-1\right)\right\}\,, \quad i=0,2,4 \,,
\eea
with the normalized $h_i^{Nor}$ defined by ${h_i}^{Nor}(N_c=\infty)=\frac{N_c}{3}{h_i}(N_c=\infty)$. 
The fitted results of $h_i$ in Table~\ref{tab:LECs} can be considered as their values at $N_C=3$. 
The quantities $h_i(N_c=\infty)$ in the previous formulae stand for their values at large $N_C$, 
which can be estimated from tree-level heavy resonance exchanges~\cite{RShis}. 
The resonance-saturation predictions of the  $h_i$ from Ref.~\cite{RShis} are 
\bea
& h_{0}(N_c=\infty)=0\ ,\,\,& h_{2}(N_c=\infty)=0\ ,\quad h_{4}'(N_c=\infty)=0\ , \nonumber \\ &
 h_1(N_c=\infty)=0.42\ , \,\, & h_{3}(N_c=\infty)=2.23\ ,\quad h_5'(N_c=\infty)=-1.45\,,
\eea
which can be compared to our Fit-6C results
\bea
&& h_{0}=0.033\ ,\,\, h_{2}=-0.08^{+0.29}_{-0.28} \ , \,\, h_4' = -0.21^{+0.29}_{-0.27} \ , \nonumber\\
&&h_1=0.43\ , \,\, h_{3}= 3.79^{+0.38}_{-0.38}\ ,\,\, h_5'=-1.78^{+0.19}_{-0.19}~\,.
\eea
As we can see, the resonance-saturation determinations are quite consistent with the Fit-6C results.

With the above preparations, we are at the point to study the influence of the subleading $N_C$ scalings, specified by Eqs.~(\ref{ncnlo1},\ref{ncnlo2}), on the  determinations of the pole trajectories with varying $N_C$. The corresponding results together with the ones by only considering the leading $N_C$ scalings, are shown in Figs.~\ref{fig:NcDs0pole} and~\ref{fig:Ncs0i12pole} for the $D^{*}_{s0}(2317)$ and the poles in the $(S,I)=(0,1/2)$ channel, respectively. Qualitatively speaking, our conclusions based on the leading $N_C$ scaling behaviors of $h_i$ are not changed when including the subleading $N_C$ effects. Taking the $D^{*}_{s0}(2317)$ trajectories for example, the only change is that the 
meeting point for the bound and virtual poles is at $N_C=8$ in the case that the subleading $N_C$ effects are included, while in the leading $N_C$ case it happens at $N_C=6$. But the trend of the pole trajectories when increasing the values of $N_C$ are almost the same in both cases. 
Similar conclusions can be also made for the poles in the $(S,I)=(0,1/2)$ channel, as can be seen from Fig.~\ref{fig:Ncs0i12pole}. 

The $1/N_C$ corrections of the subtraction constant  $a$ contributes another source of subleading $N_C$ scaling effects. 
Since the subtraction constant is introduced through the unitarization procedure, it is rather difficult 
to directly estimate its leading $N_C$ value, in contrast to the $h_i$ situation. We simply vary the values of 
the scale $\mu$ in Eq.~\eqref{eq:g}, but keeping the subtraction constant $a$ fixed at its fitted value, to roughly  estimate the subleading $N_C$ scaling effects of $a$. Several different values of the $\mu$ from 0.9~GeV to 1.2~GeV are used to study the pole trajectories for the $D^{*}_{s0}(2317)$ and the $(S,I)=(0,1/2)$ channel. Since we do not observe qualitative changes by using different values of $\mu$ and they look similar as the cases by introducing the subleading $N_C$ scaling effects in the $h_i$, we shall not show explicitly these plots in order to avoid overloading  the manuscript with too many figures.}



\begin{figure}[ht]
   \centering
   \includegraphics[width=0.9\textwidth]{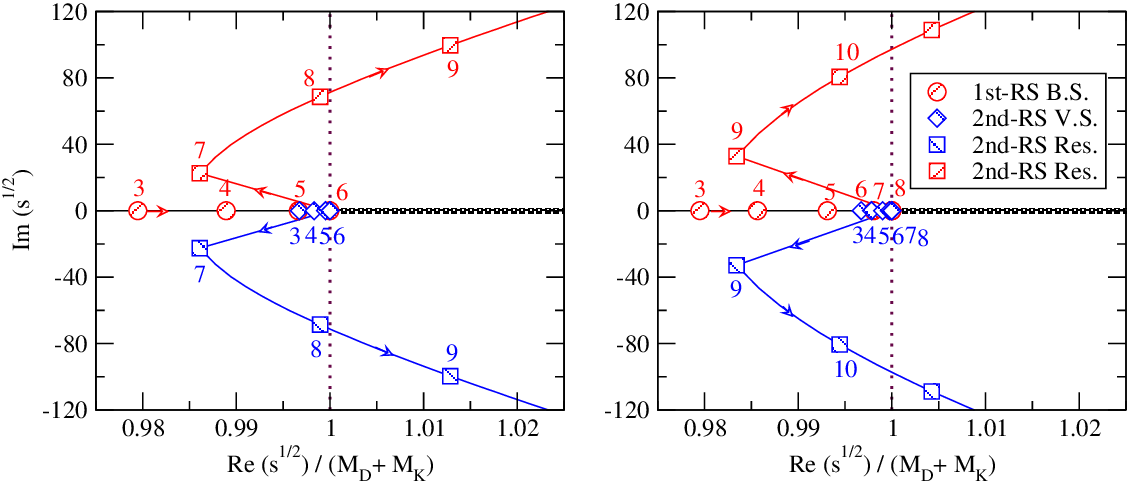} 
  \caption{$N_C$ trajectories of the $\ds$ pole. The numbers in the plots indicate the values of $N_C$. Left: results with leading $N_C$ scaling laws in Eq.~\eqref{Eq:hisNc}; Right: results by including the subleading $N_C$ scaling effects in Eqs.~\eqref{ncnlo1} and \eqref{ncnlo2}. }
   \label{fig:NcDs0pole}
\end{figure}

\begin{figure}[ht]
   \centering
   \includegraphics[width=0.9\textwidth]{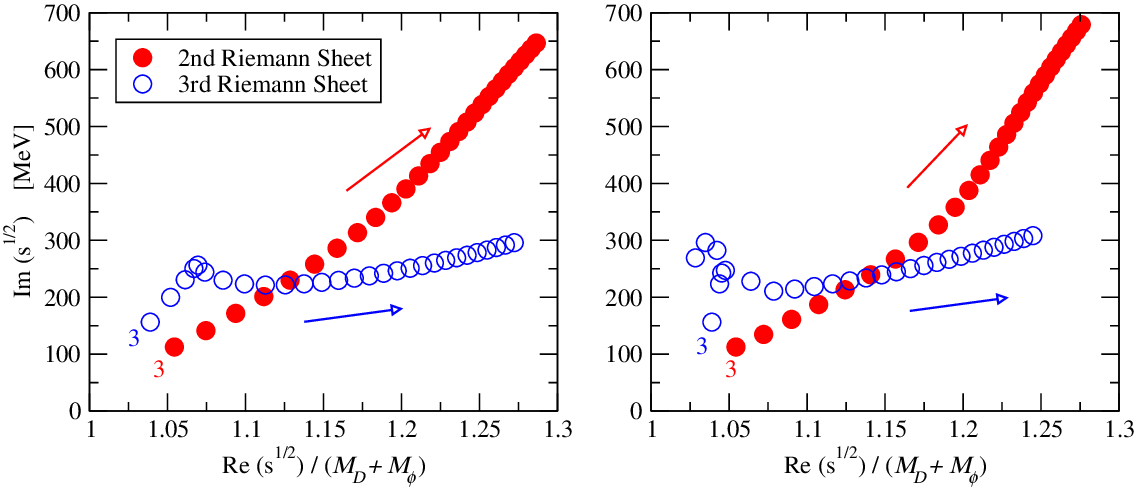} 
  \caption{$N_c$ trajectories of the poles in the channel $(0,1/2)$ with $N_C$ 
           varying from 3 to 30. The symbol $\phi$ in the horizontal-axis label 
           is $\pi$ for the 2nd RS, and $\eta$ for the 3rd RS. Left: results with leading $N_C$ scaling laws in Eq.~\eqref{Eq:hisNc}; Right: results by including the subleading $N_C$ scaling effects in Eqs.~\eqref{ncnlo1} and \eqref{ncnlo2}.   }
   \label{fig:Ncs0i12pole}
\end{figure}

\section{Conclusions}\label{sect.concl}

In this work we have calculated the scattering of pNGBs ($\pi,K,\eta,\eta'$) 
off charmed meson ($D$ or $D_s$) and then unitarized the perturbative 
chiral amplitudes to investigate the charm scalar mesons. Recent lattice 
simulation data have been used in our study to constrain the free parameters 
in the fits.  In addition to the prediction of the $D\pi$ scattering lengths 
with varying pion mass in the $(S,I)=(0,1/2)$ channel, we also give predictions 
for all of the relevant scattering lengths with physical masses.  
Careful and extensive analyses of the charm scalar poles and their residues 
in different $(S,I)$ channels have been carried out, which are summarized in 
Table~\ref{tab:pole6c}. A virtual pole is found to be responsible for the 
positive scattering length in the $(S,I)=(-1,0)$ channel. Two 
poles are observed for the $(S,I)=(1,1)$ case: the one on the second sheet 
may explain the enhancement recently reported in Ref.~\cite{Purohit:2015zka}. 
The pole positions for the $\ds$ and the channel with $(S,I)=(0,1/2)$ are 
compatible with previous determinations, with the exception that 
our determination for the $(S,I)=(0,1/2)$ pole at around 2.4~GeV has larger width, 
which is close to the PDG value. 
Both $M_\pi$ and $N_C$ trajectories for the $\ds$ and the two poles in the
 $(S,I)=(0,1/2)$ channel are 
studied in detail. The $M_\pi$ trajectories of the pole around 2.1~GeV quite 
resembles the ones of the $f_0(500)$ obtained in 
Refs.~\cite{Hanhart:2008mx,Hanhart:2014ssa}. 
The $N_C$ trajectories of the $\ds$ show that this physical bound state 
becomes a resonance for $N_C>6$. For large values of $N_C$, the trajectories of 
the $\ds$ and the poles with $(S,I)=(0,1/2)$ do not tend to fall down to the 
real axis, indicating that they do not behave like the standard quark-antiquark 
mesons of QCD.

\section*{Acknowledgements}

We thank Dr.~Feng-Kun Guo for useful discussions. 
This work was carried out in the framework of the Sino-German Collaborative
Research Center ``Symmetries and the Emergence of Structure in QCD'' (CRC~110)
co-funded by the DFG and the NSFC.
This work is supported in part by the National Natural Science Foundation of China (NSFC) under Grant Nos.~11575052 and 11105038, the Natural Science Foundation of Hebei Province with contract No.~A2015205205, 
the grants from the Education Department of Hebei Province under contract No.~YQ2014034, 
the grants from the Department of Human Resources and Social Security of Hebei Province with contract No.~C201400323, 
the Doctor Foundation of Hebei Normal University under Contract No.~L2010B04,
and by the Chinese Academy of Sciences (CAS) President's International Fellowship Initiative (PIFI) (Grant No.
2015VMA076).

\section*{Appendix: Some of the coefficients in Table~\ref{tab:ci} }\label{app1}

Here, we give the  form of the coefficients in Table~\ref{tab:ci} that were not explicitely stated there:   
\begin{eqnarray}
 C_1^{1,0 \,\, D K\to D_s\eta} &=& \frac{-M_K^2(5c_\theta+4\sqrt2 s_\theta)+3M_\pi^2 c_\theta }{2\sqrt3}\,,  \\
 C_{35}^{1,0 \,\, D K\to D_s\eta} &=& \frac{c_\theta + 2\sqrt2 s_\theta}{\sqrt{3}} \,,   
 \end{eqnarray}

\begin{eqnarray}
 C_0^{1,0 \,\, D_s\eta\to D_s\eta} &=& \frac{c_\theta^2(4M_K^2-M_\pi^2) + 4\sqrt2 c_\theta s_\theta (M_K^2-M_\pi^2)+ s_\theta^2(2M_K^2+M_\pi^2)}{3} \,,\\
 C_1^{1,0 \,\, D_s\eta\to D_s\eta} &=& \frac{2(M_\pi^2-2M_K^2)  (\sqrt2 c_\theta + s_\theta)^2 }{3} \,, \\
 C_{35}^{1,0 \,\, D_s\eta\to D_s\eta} &=& \frac{ 2 (\sqrt2 c_\theta + s_\theta)^2 }{3} \,, 
 \end{eqnarray} 
 
\begin{eqnarray}
 C_1^{1,0 \,\, D K\to D_s\eta'} &=& \frac{  M_K^2(4\sqrt2 c_\theta-5s_\theta) +3M_\pi^2 s_\theta }{2\sqrt3} \,, \\
 C_{35}^{1,0 \,\, D K\to D_s\eta'} &=&  \frac{s_\theta-2\sqrt2 c_\theta}{\sqrt3} \,, 
 \end{eqnarray} 

\begin{eqnarray}
 C_0^{1,0 \,\, D_s\eta\to D_s\eta'} &=& \frac{2(M_\pi^2-M_K^2)(\sqrt2 c_\theta^2 - c_\theta s_\theta - \sqrt2 s_\theta^2) }{3} \,,\\
 C_1^{1,0 \,\, D_s\eta\to D_s\eta'} &=& \frac{2(2M_K^2-M_\pi^2)(\sqrt2 c_\theta^2 - c_\theta s_\theta - \sqrt2 s_\theta^2) }{3} \,,  \\
 C_{35}^{1,0 \,\, D_s\eta\to D_s\eta'} &=& \frac{ -2 (\sqrt2 c_\theta^2 - c_\theta s_\theta - \sqrt2 s_\theta^2) }{3} \,, 
 \end{eqnarray} 

\begin{eqnarray}
 C_0^{1,0 \,\, D_s\eta'\to D_s\eta'} &=& \frac{s_\theta^2(4M_K^2-M_\pi^2) + 4\sqrt2 c_\theta s_\theta (M_\pi^2-M_K^2)+ c_\theta^2(2M_K^2+M_\pi^2)}{3} \,,\\
 C_1^{1,0 \,\, D_s\eta'\to D_s\eta'} &=& \frac{2(M_\pi^2-2M_K^2)  (\sqrt2 s_\theta - c_\theta)^2 }{3} \,, \\
 C_{35}^{1,0 \,\, D_s\eta'\to D_s\eta'} &=& \frac{ 2 (\sqrt2 s_\theta - c_\theta)^2 }{3} \,, 
 \end{eqnarray}

\begin{eqnarray}
 C_{0}^{0,\frac12\,\,D\eta\to D\eta}&=& \frac{c_\theta^2(4M_K^2-M_\pi^2) + 4\sqrt2 c_\theta s_\theta (M_K^2-M_\pi^2)+ s_\theta^2(2M_K^2+M_\pi^2)}{3} \,,\\
 C_{1}^{0,\frac12\,\,D\eta\to D\eta} &=& \frac{-M_\pi^2 (\sqrt2 s_\theta - c_\theta)^2 }{3} \,, \\
 C_{35}^{0,\frac12\,\,D\eta\to D\eta} &=& \frac{ (\sqrt2 s_\theta - c_\theta)^2 }{3} \,, 
 \end{eqnarray} 
 
\begin{eqnarray}
 C_{1}^{0,\frac12\,\,D_s K\to D\eta} &=& \frac{c_\theta (5M_K^2-3M_\pi^2) + 4\sqrt2 s_\theta M_K^2 }{2\sqrt6} \,,\\
 C_{35}^{0,\frac12\,\,D_s K\to D\eta} &=& \frac{-(2\sqrt2 s_\theta + c_\theta) }{\sqrt6} \,, 
 \end{eqnarray}

\begin{eqnarray}
 C_{0}^{0,\frac12\,\,D\eta\to D\eta'}&=& \frac{ 2(M_\pi^2-M_K^2) (\sqrt2 c_\theta^2 - c_\theta s_\theta - \sqrt2 s_\theta^2) }{3} \,,\\
 C_{1}^{0,\frac12\,\,D\eta\to D\eta'} &=& \frac{ M_\pi^2 (-\sqrt2 c_\theta^2+  c_\theta s_\theta + \sqrt2 s_\theta^2)}{3} \,, \\
 C_{35}^{0,\frac12\,\,D\eta\to D\eta'} &=& \frac{  (\sqrt2 c_\theta^2 - c_\theta s_\theta - \sqrt2 s_\theta^2) }{3} \,, 
 \end{eqnarray}

\begin{eqnarray}
  C_{1}^{0,\frac12\,\,D_s\bar K\to D\eta'}&=& \frac{ (5M_K^2- 3M_\pi^2) s_\theta - 4\sqrt2 M_K^2 c_\theta }{2\sqrt6} \,, \\
 C_{35}^{0,\frac12\,\,D_s\bar K\to D\eta'} &=& \frac{ 2\sqrt2 c_\theta -  s_\theta }{\sqrt6} \,, 
 \end{eqnarray}

\begin{eqnarray}
 C_0^{1,0 \,\, D\eta'\to D\eta'} &=& \frac{s_\theta^2(4M_K^2-M_\pi^2) + 4\sqrt2 c_\theta s_\theta (M_\pi^2-M_K^2)+ c_\theta^2(2M_K^2+M_\pi^2)}{3} \,,\\
 C_1^{1,0 \,\, D\eta'\to D\eta'} &=& \frac{-M_\pi^2(\sqrt2 c_\theta + s_\theta)^2 }{3} \,, \\
 C_{35}^{1,0 \,\, D\eta'\to D\eta'} &=& \frac{ (\sqrt2 c_\theta + s_\theta)^2 }{3} \,.
 \end{eqnarray}

\end{document}